\documentclass{svproc}
\usepackage{amsmath}
\usepackage{amsfonts}
\usepackage{numcompress}
\usepackage[numbers, sort&compress]{natbib}
\usepackage{amssymb}
\usepackage{graphicx}
\usepackage{lmodern}
\usepackage{hyperref}
\usepackage{dcolumn}
\usepackage{bm}
\usepackage[mathlines]{lineno}
\usepackage{float}
\usepackage[T1]{fontenc}
\usepackage{subfigure}
\usepackage{url}

\DeclareUnicodeCharacter{2212}{\ensuremath{-}}
\begin{document}
\mainmatter              
\title{Equal Majorana Phases from a Minimal and Predictive Neutrino Texture}

\author{Sagar Tirtha Goswami\inst{1}, Pralay Chakraborty\inst{1}\and Subhankar Roy\inst{1}
}
\authorrunning{Sagar Tirtha Goswami et al.} 
%
\tocauthor{Subhankar Roy}
\institute{Department of Physics, Gauhati University,\\
\email{sagartirtha@gauhati.ac.in, pralay@gauhati.ac.in, subhankar@gauhati.ac.in}
}

\maketitle              

\begin{abstract}
We propose a new, minimal Majorana neutrino mass matrix texture and study its predictions within a partial tri-bimaximal (TBM) mixing scheme, where \(\sin\theta_{12} = 1/\sqrt{3}\) and \(\sin\theta_{23} = 1/\sqrt{2}\), with \(\theta_{13}\) and \(\delta\) treated as free parameters. The texture forbids \(\theta_{13} = 0\) and does not correspond to a \(\mu\text{–}\tau\) symmetric structure. As a notable feature, the texture predicts the equality of the Majorana phases. In addition, we find that the predictions show distinct behavior depending on the sign of a single real texture parameter.  The texture is realized through a hybrid framework of one Type-I seesaw and two Type-II seesaw mechanisms under an extended symmetry group, $SU(2)_L \otimes U(1)_Y \otimes A_4 \otimes Z_{10} \otimes Z_{7}$ with properly chosen model parameters. The texture is seen to favor the normal hierarchy for neutrino masses.
\end{abstract}
 
\newcolumntype{P}[1]{>{\centering\arraybackslash}p{#1}}

\section{Introduction}
\label{introduction}

In the realm of particle physics,  neutrino $(\nu)$ is the most evasive particle.  The standard model (SM),  which is quite successful in describing the properties of other particles, has proved to be inadequate in the neutrino sector.  In particular, the SM is silent on two key issues: the origin of tiny but non-zero masses of neutrinos, confirmed by various neutrino oscillation experiments\,\cite{Davis:1968cp, SNO:2001kpb, Super-Kamiokande:2001ljr, Bionta:1987qt, Super-Kamiokande:1998kpq, KamLAND:2002uet, K2K:2002icj, T2K:2013ppw, Kamiokande-II:1989hkh}, and the significant mixing among different flavours of neutrinos\,\cite{Esteban:2024eli,ParticleDataGroup:2022pth}. The quest to find answers to these unresolved questions among many others has led us beyond the SM (BSM), which is an arena of various models and conjectures.  These models try to predict some observational parameters related to neutrinos,  such as the mass squared differences\,($\Delta m^2_{21}$ and $\Delta m^2_{31}$),  three mixing angles ($\theta_{12}$,\,$\theta_{23}$ and $\theta_{13}$), one Dirac CP-violating phase ($\delta$) and two Majorana CP-violating phases ($\alpha$ and $\beta$). So far, the experiments have determined the mass squared differences and the mixing angles with great precision\,\cite{ParticleDataGroup:2022pth}.  But the octant of $\theta_{23}$, and precise range of $\delta$ are still undetermined. In addition, the individual mass eigenvalues\,($m_1$,\,$m_2$ and $m_3$) and Majorana phases ($\alpha$,\,$\beta$) are not witnessed by oscillation experiments.

The information about all the parameters mentioned above is contained in the Majorana neutrino mass matrix ($M_\nu$). 
Although a complete quantitative determination of the observable parameters remains unavailable,  even the limited information already provides valuable insights into the underlying symmetries governing neutrino mass and mixing patterns.
Based on this knowledge, the theorists, in general, try to obtain the $M_\nu$ from first principle,  which predict not only the observable parameters already determined,  but also the ones yet to be measured. However, as we wait to see experimental verification of these predictions, certainly avenues are open for the proposition of a new $M_\nu$.

As a common practice, some phenomenological ideas such as $\mu-\tau$ symmetry\,\cite{King:2018kka,Xing:2015fdg,Altarelli:2005yx}, texture zeroes\,\cite{Ludl:2014axa}, hybrid textures\,\cite{Kalita:2015tda,Singh:2018bap}, mixed $\mu-\tau$ symmetry\,\cite{Dey:2022qpu, Chakraborty:2023msb} etc., are assigned to $M_\nu$ to expect certain results of the observational parameters. These results, which can be called mixing schemes, basically involve the mixing angles and the Dirac CP phase. Over the time, different mixing schemes have been postulated, which are related to specific phenomenological forms of $M_\nu$. For example, if we talk about $\mu-\tau$ symmetry, we see that the mixing scheme predicted by such a texture always has $\theta_{23}=45^\circ$ and $\theta_{13}=0$, while $\theta_{12}$ remains free and is related to texture parameters. A careful calibration of the texture parameters can make $\sin\,\theta_{12}=1/\sqrt{3}$ and this particular mixing scheme is known as TBM mixing\,\cite{Harrison:2002er}.

However, it is important to note that the value of $\theta_{13}$ is found to be non-zero, which contradicts the original TBM scheme and is inconsistent with the prediction of a $\mu-\tau$ symmetric texture.  Interestingly the other two angles, $\theta_{12}$ and $\theta_{23}$, still lie within the $3\sigma$ range of present experimental data\,\cite{Esteban:2024eli,ParticleDataGroup:2022pth}. This renders the TBM scheme only partially viable, while ruling out the strict realization of $\mu-\tau$ symmetry. In addition, the said symmetry stays silent on the predictions of two Majorana phases, two important parameters in case neutrinos were proven to be Majorana particles.

The partial validity of TBM scheme opens two interesting possibilities. One is based on a top down approach in which new textures are put forward from first principle to predict a non zero $\theta_{13}$, while resembling the TBM scheme in other aspects. A common way in this regard is the addition of corrections to a $\mu-\tau$ symmetric texture through various means\,\cite{King:2002nf,Frampton:2004ud,Altarelli:2004jb,Ganguly:2022qxj,Borah:2019ldn,Rodejohann:2004cg,Borah:2013jia, King:2011zj,Altarelli:2005yx,Honda:2008rs}. The other one is based on a bottom up approach in which one can choose a valid partial TBM scheme and apply it to a texture to study its phenomenology\,\cite{Shimizu:2011xg, Xing:2002sw,Xing:2006ms,Adhikary:2006jx,King:2007pr,Brahmachari:2008fn,Adhikary:2008au,Hirsch:2008mg,Morisi:2009qa,Hayakawa:2009va,Goswami:2009yy,Barry:2010zk,Albright:2010ap,Ishimori:2010fs,King:2011ab,Antusch:2011qg,Ahn:2012tv,Ishimori:2012fg,Rodejohann:2012cf,Hagedorn:2012ut,King:2013eh}.

In this work, we take a bottom up approach in which we adopt a partial TBM scheme. But rather than fixing $\theta_{13}$ with a single value, we keep it as a free parameter. We also posit a new minimal Majorana mass matrix texture with four complex parameters and try to see its various predictions including the Majorana phases under the chosen mixing scheme. 

The plan of the paper is as follows: In Section\,\ref{sec2}, we discuss the formalism of our work, while in Section\,\ref{sec3}, we discuss the results of our texture under various conditions. In Section\,\ref{sec5}, a symmetry perspective of our proposed texture is discussed. Finally we give a summary of our work in Section\,\ref{sec6}.

\section{Formalism \label{sec2}}
We begin by outlining the theoretical framework used in this work. This includes the structure of the proposed Majorana mass matrix texture and the parametrization of the leptonic mixing matrix under the partial TBM scheme ($\sin\,\theta_{12}=1/\sqrt{3}$,\,$\sin\,\theta_{23}=1/\sqrt{2}$, while $\theta_{13}$ and $\delta$ being free parameters), as discussed in Section\,\ref{introduction} .

The Majorana mass matrix texture proposed in this work is given by,

\begin{equation}
\label{Mnu}
M_\nu = \begin{bmatrix}
a+\frac{9}{5}h & a+b & -a+b \\
 a+b & a+g+h & -a+g\\
 -a+b & -a +g & a+g-h
\end{bmatrix},
\end{equation}
 where the four parameters, $a,\,b,\,g,$ and $h$ are complex in general.

 To obtain the information of the physical parameters, we need to diagonalize $M_\nu$ with Pontecorvo-Maki-Nakagawa-Sakata (PMNS) matrix, $U$\,\cite{Maki:1962mu}, as shown below,
 \begin{equation}
M_\nu^{diag}=U^T M_\nu U,\nonumber
\end{equation}
 
 where
 
 \begin{equation}
U = \begin{bmatrix}
e^{i\phi_1} & 0 & 0\\
0 & e^{i\phi_2} & 0\\
0 & 0 & e^{i\phi_3}
\end{bmatrix} \times \begin{bmatrix}
1 & 0 & 0\\
0 & c_{23} & s_{23}\\
0 & - s_{23} & c_{23}
\end{bmatrix}\times \begin{bmatrix}
c_{13} & 0 & -s_{13}\,e^{-i\delta}\\
0 & 1 & 0\\
s_{13} e^{i\delta} & 0 & c_{13}
\end{bmatrix} \times\begin{bmatrix}
c_{12} & s_{12} & 0\\
-s_{12} & c_{12} & 0\\
0 & 0 & 1
\end{bmatrix},
\label{U}
\end{equation}

with $s_{ij}=\sin\theta_{ij}$ and $c_{ij}=\cos\theta_{ij}$.  It is to be mentioned that the unphysical phases $\phi_1$, $\phi_2$ and $\phi_3$ can be absorbed by the redefinition  of the charged lepton fields in terms of these phases.  In general, the Majorana phases $\alpha$ and $\beta$ are embedded into $U$.  On the other hand,  these phases can also be placed inside the diagonal neutrino mass matrix.  The present work adopts the latter approach as shown below,

\begin{eqnarray}
M_\nu^{\text{diag}} &=& \begin{bmatrix}
m_1 e^{-2i\alpha} & 0 & 0\\
0 & m_2 e^{-2i\beta} & 0\\
0 & 0 & m_3
\end{bmatrix}. \nonumber
\end{eqnarray}


The advantage of the above mentioned parametrization of  $M_\nu^{\text{diag}}$ is that the Majorana phases $\alpha$ and $\beta$ can be directly obtained from the following relations,

\begin{eqnarray}
\alpha&=&-\frac{1}{2} Arg\,[M^{diag}_{\nu_{11}}] \label{alpha eq},\nonumber \\
\beta&=&-\frac{1}{2} Arg\,[M^{diag}_{\nu_{22}}] \label{beta eq}.
\end{eqnarray}

The present analysis involves four complex texture parameters and nine observational parameters. We have three complex diagonalizing conditions, viz., $M_{\nu_{12}}^{\text{diag}}=0,\,M_{\nu_{13}}^{\text{diag}}=0,\,M_{\nu_{23}}^{\text{diag}}=0$, appearing as transcendental equations. Additionally, the chosen parametrization ensures that $M_{\nu_{33}}^{diag}$ is real. Therefore, if we exploit these conditions,  some of the free parameters can be reduced. 

Now, if we apply our chosen partial TBM scheme to $U$ (Eq.\,\ref{U}), we find the following three relations from the diagonalization conditions,
\begin{eqnarray}
\label{a}
a&=&\frac{h}{\sqrt{2}}\cot\,\theta_{13}\,e^{i\delta},\\
\label{b}
b&=&\frac{h}{2}\,\cot\,\theta_{13}\,\left[-\frac{9\sqrt{2}}{5}\,e^{-i\delta}-\tan\,\frac{\theta_{13}}{2}\right],\\
\label{c}
g&=&\frac{h\,e^{-2i\delta}}{8}\,\left[\frac{36\cot^2\,\theta_{13}}{5}+\frac{\sqrt{2}\,e^{i\delta}\,\csc\,\theta_{13}}{1+\cos\,\theta_{13}}\,(1
+2\cos\,\theta_{13} +3\cos\,2\theta_{13})\right].
\end{eqnarray}
Further, on applying the constraint that $M_{\nu_{33}}^{diag}$ is real, we reduce another parameter,  $\text{Im}\,[h]$ as follows,

\begin{equation}
\label{imh}
\text{Im}\,[h]=\frac{\text{Re}\,[h] \sin\,\delta\,\left[72 \cos\,\delta\,\cos\,\left(\frac{\theta_{13}}{2}\right)+5\sqrt{2}\,\left(\sin\,\left(\frac{\theta_{13}}{2}\right)+\sin\,\left(\frac{5\theta_{13}}{2}\right)\right)\right]}{36\cos\,\left(2\delta\right)\cos\,\left(\frac{\theta_{13}}{2}\right)+5\sqrt{2}\cos\,\delta\,\left(\sin\,\left(\frac{\theta_{13}}{2}\right)+\sin\,\left(\frac{5\theta_{13}}{2}\right)\right)}. 
\end{equation}

So, from Eqs.\,(\ref{a})-(\ref{imh}), we can say that there are three free parameters, viz., $\text{Re}[h]$, \,$\theta_{13}$ and $\delta$, that determine the predictions of the texture (Eq.\,\ref{Mnu}).  Again, from Eq.\,(\ref{a}), we can relate $\theta_{13}$ and $\delta$ to texture parameters via two simple relations as shown in the following,
\begin{eqnarray}
\label{equation:2}
\sin^2\theta_{13}  &=& \frac{1}{1+2\,|\frac{a}{h}|^2},
\quad
\delta = \text{Arg} \left[\frac{a}{h}\right]. \nonumber
\end{eqnarray}

From Eq.\,(\ref{a}), it is easy to see that if $\theta_{13}=0$, the three texture parameters, $a,\,b,\,\text{and}\, g$ become undefined. Hence, our proposed texture (Eq.\,(\ref{Mnu})) strictly forbids a vanishing $\theta_{13}$ under the chosen mixing scheme.  Likewise, we can not have $h=0$, as it would collapse the whole texture. Hence, the proposed texture is not simply a $\mu-\tau$ deviated texture.

In addition to the texture relations (Eqs.\,(\ref{a})-(\ref{imh})), we now express the analytical forms of the two Majorana phases in terms of $\text{Re}[h],\,\theta_{13}$ and $\delta$ in the following (Eq.\,(\ref{Maj})). It is notable that the two phases come out to be equal. 

\begin{align}
\label{Maj}
\alpha =& \beta =\frac{1}{2}\,\tan^{-1}\left[\frac{18\sqrt{2}\,\cos\,\left(\frac{\theta_{13}}{2}\right)\,\sin\,3\delta+5\sin\,2\delta \,\left(\sin\,\left(\frac{\theta_{13}}{2}\right)+\sin\,\left(\frac{5\theta_{13}}{2}\right)\right)}{18\sqrt{2}\,\cos\,\left(\frac{\theta_{13}}{2}\right)\,\cos\,3\delta+5\cos\,2\delta\,\left(\sin\,\left(\frac{\theta_{13}}{2}\right)+\sin\,\left(\frac{5\theta_{13}}{2}\right)\right)}\right].
\end{align}

Further, the expressions for the mass eigenvalues are given as follows,
\begin{align}
\label{phyob}
m_1=&\frac{1}{\left[36 \cos\,2 \delta\,  \cos\, (\frac{\theta_{13} }{2})+5 \sqrt{2} \cos\, \delta  \left(\sin\, (\frac{\theta_{13} }{2})+\sin \,(\frac{5 \theta_{13} }{2})\right)\right]}[(\text{Re}[h])^2 \sin ^2\,\theta_{13} \nonumber\\ 
&\sec ^4\,\left(\frac{\theta_{13} }{2}\right) (349 
+90 \sqrt{2}  \cos\,\delta\,( \,\sin\,\theta_{13}+\sin\,2 \theta_{13} +\sin\,3 \theta_{13})+\frac{623\, \cos\,\theta_{13} }{2}\nonumber \\
&+25 \cos 2 \theta_{13} 
-25 \cos\,3 \theta_{13} -\frac{25}{2} \cos\,5 \theta_{13} )]^{1/2},\nonumber 
\end{align}
\begin{align}
m_2=&\frac{1}{\sqrt{2}[20\sqrt{2}\cos\,\delta\cos ^2(\frac{\theta_{13}}{2})\,(1+\cos\,\theta_{13}+\cos\,2\theta_{13})\sin ^4(\frac{\theta_{13}}{2})+9\cos\,2\delta\sin ^3\,\theta_{13}]}\nonumber \\
&[(698 +623\cos\,\theta_{13}
+50\cos\,2\theta_{13}-50\cos\,3\theta_{13}-25\cos\,5\theta_{13}+180\sqrt{2}\cos\,\delta \nonumber \\
&(\sin\,\theta_{13} 
+\sin\,2\theta_{13}+\sin\,3\theta_{13}))(\text{Re}[h])^2\cos^2
\left(\frac{\theta_{13}}{2}\right)\sin^4\left(\frac{\theta_{13}}{2}\right)(1+2\cos\,\theta_{13})^2 ]^{\frac{1}{2}},\nonumber \\
m_3=&\frac{1}{20\left[5\sqrt{2}\,\cos\,\delta\,\left(1+\cos\,\theta_{13}+\cos\,2\theta_{13}\right)+180\,\cos\,2\delta\,\cot\,(\frac{\theta_{13}}{2})\right]}\,[360\sqrt{2}\cos\,\delta \nonumber \\
&
(1
+\cos\,\theta_{13}+\cos\,2\theta_{13})
+623 \cot\,\theta_{13}+(698+50\cos\,2\theta_{13}-50\cos\,3\theta_{13}
- \nonumber \\
&25\cos\,5\theta_{13})\csc\,\theta_{13}]\,\text{Re}[h]\,\csc ^2\,\theta_{13}.
\end{align}

In the next section, we will try to visualize the numerical ranges of the physical parameters.

\section{Numerical Analysis \label{sec3}}
In this section, we will try to observe the patterns of the physical parameters (Eq.\,(\ref{phyob})), and extract their numerical bounds. For that, we take $3\sigma$ ranges\,\cite{Esteban:2024eli} of two of the input parameters, $\theta_{13}$ and $\delta$, while varying $\text{Re}[h]$ randomly in the interval $[-1,1]$. Interestingly, two distinct scenarios for the features of the observable parameters emerge, depending on the sign of $\text{Re}[h]$.
We categorise the results into two cases as shown below,\\

\subsection*{Case I}

\begin{figure}
  \centering
  \subfigure[]{\includegraphics[width=0.48\textwidth]{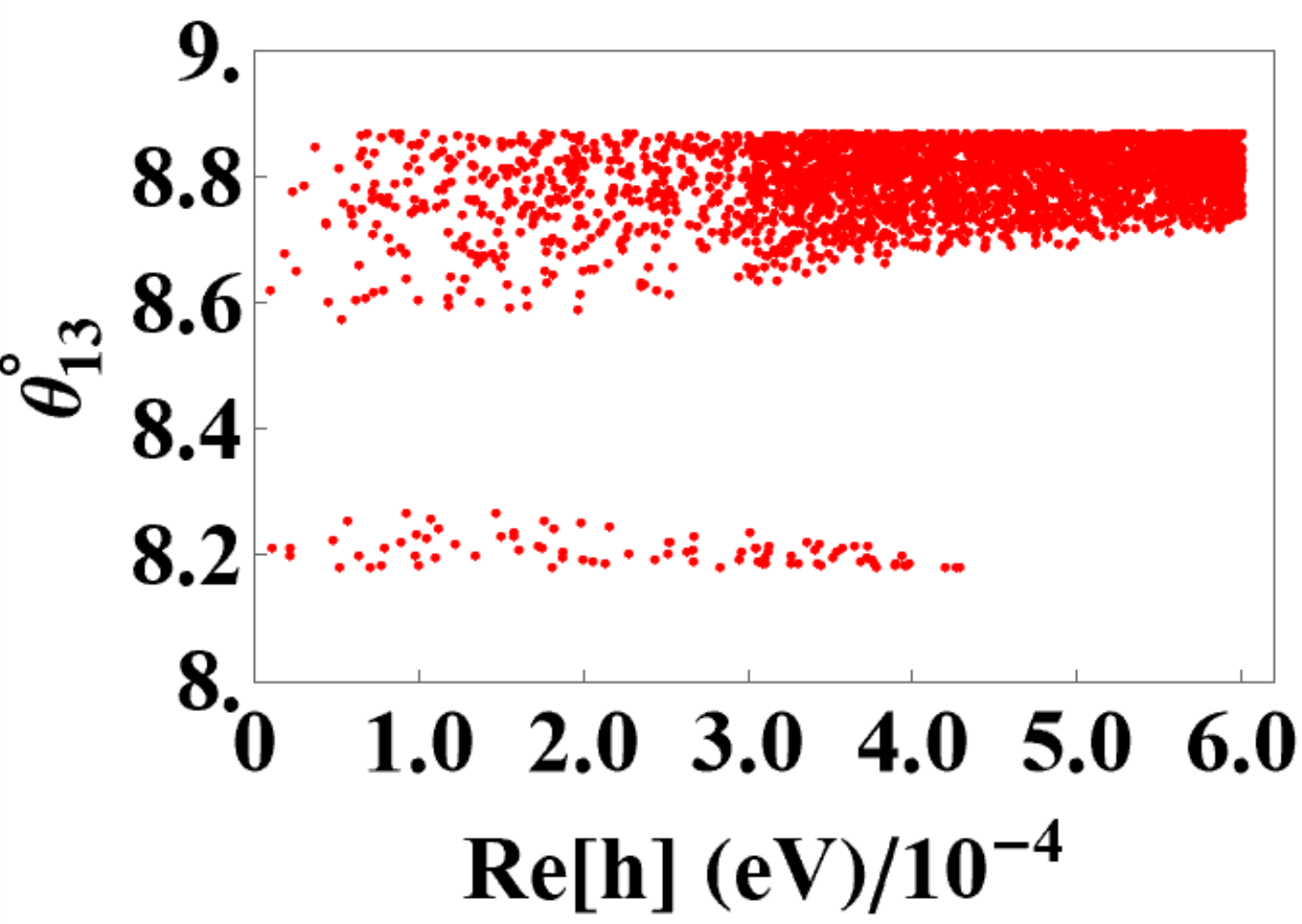}\label{fig:1(a)}}
  \subfigure[]{\includegraphics[width=0.48\textwidth]{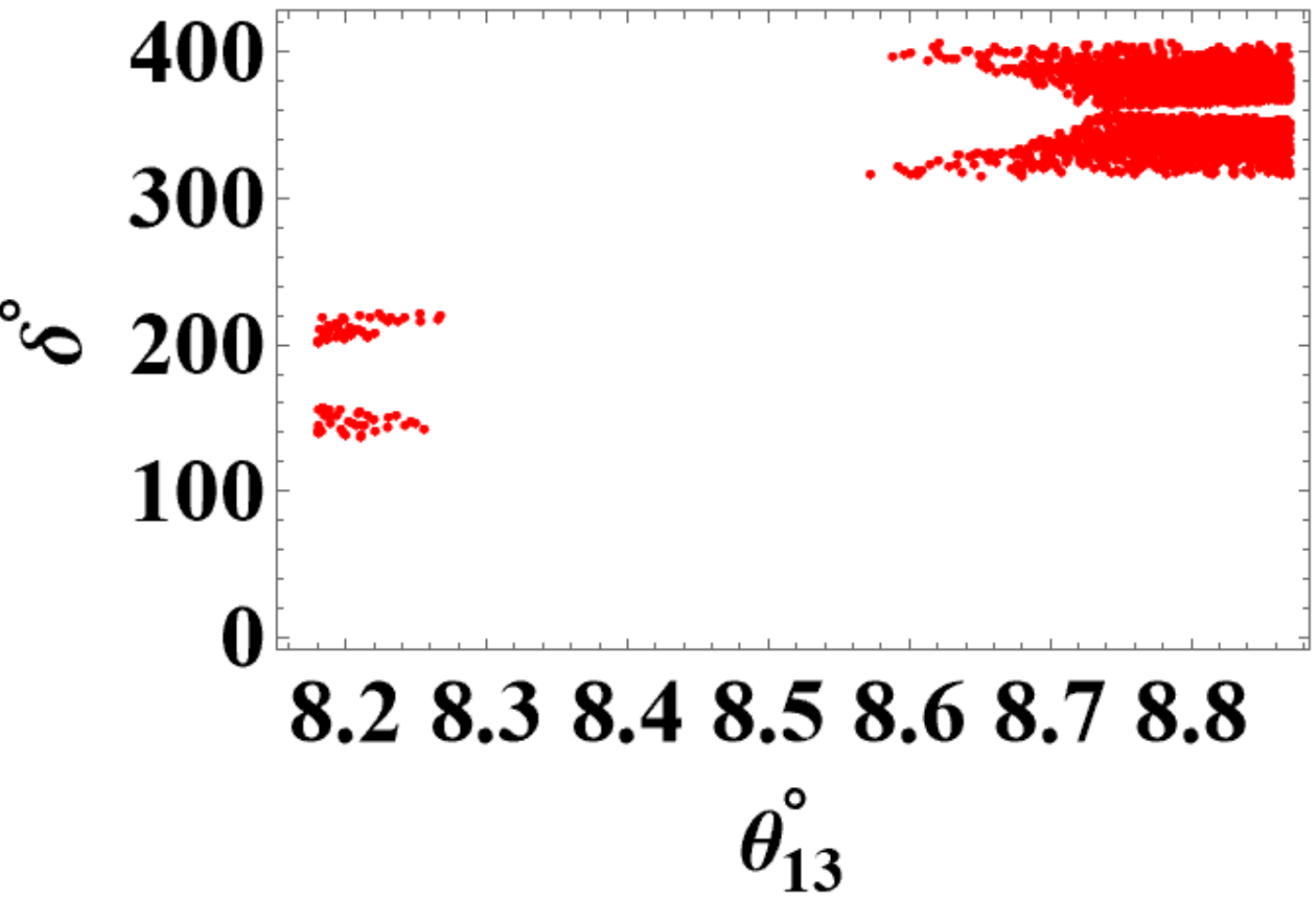}\label{fig:1(b)}}
   \subfigure[]{\includegraphics[width=0.48\textwidth]{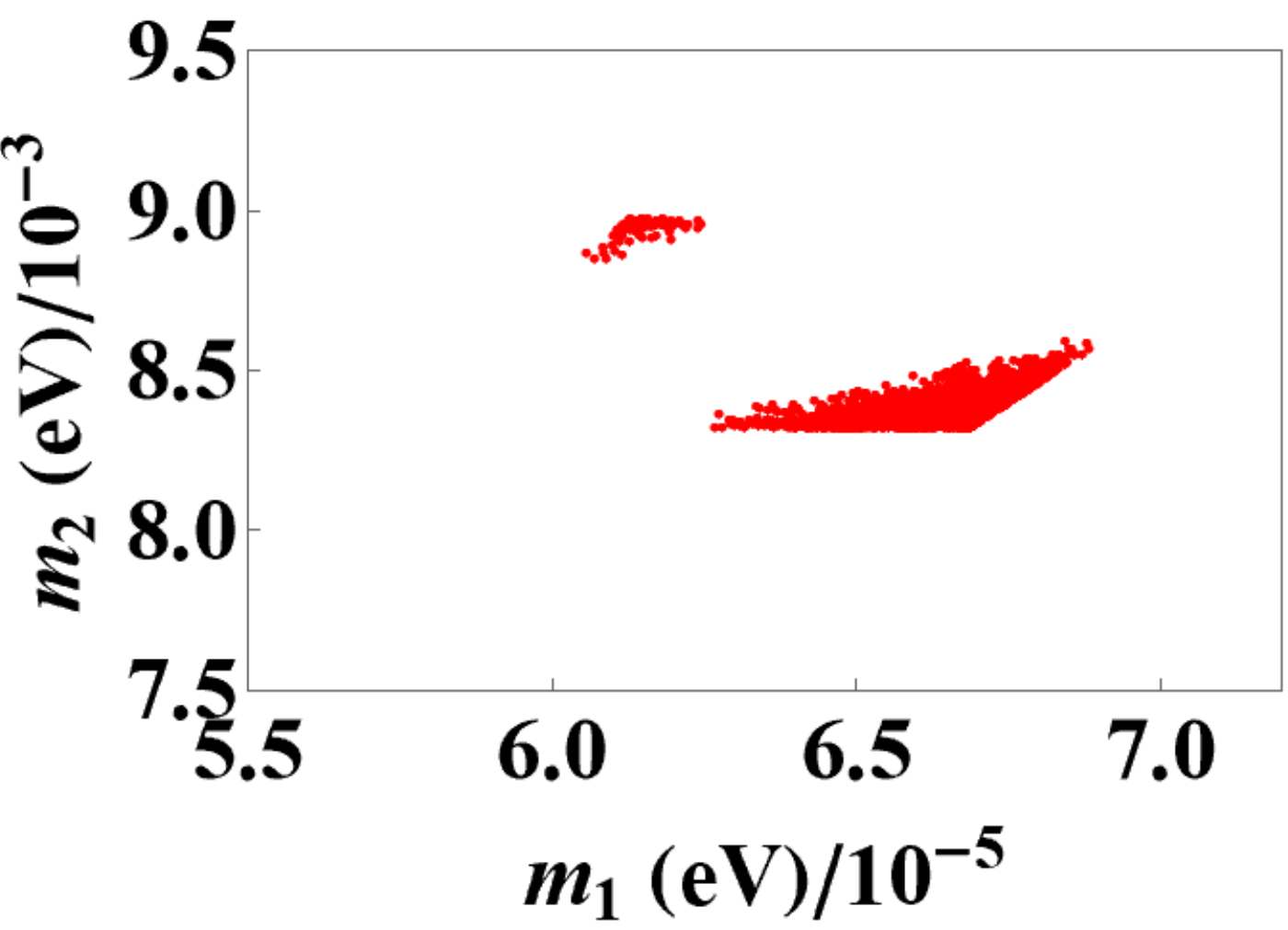}\label{fig:1(c)}} 
    \subfigure[]{\includegraphics[width=0.48\textwidth]{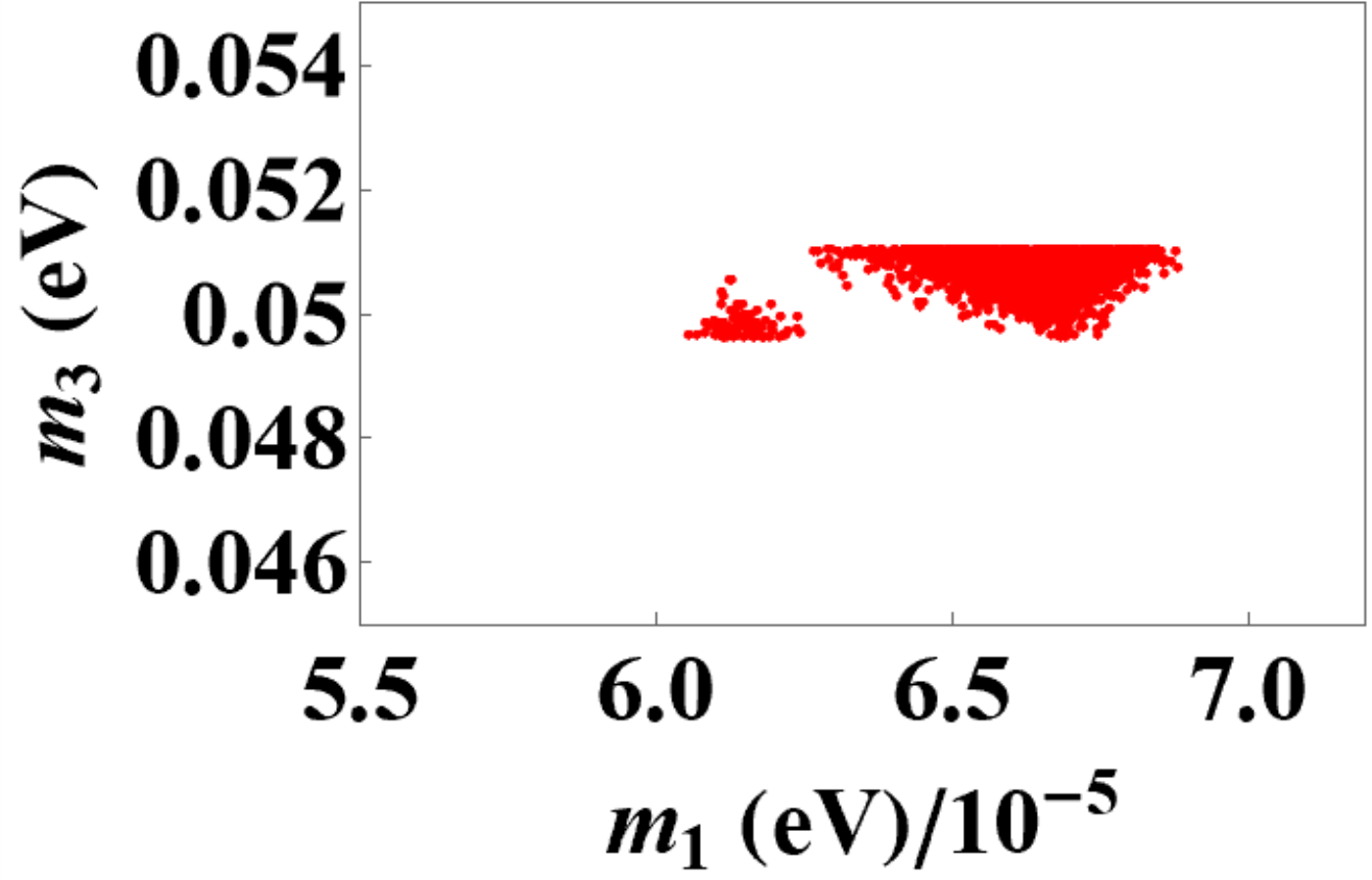}\label{fig:1(d)}}
     \subfigure[]{\includegraphics[width=0.48\textwidth]{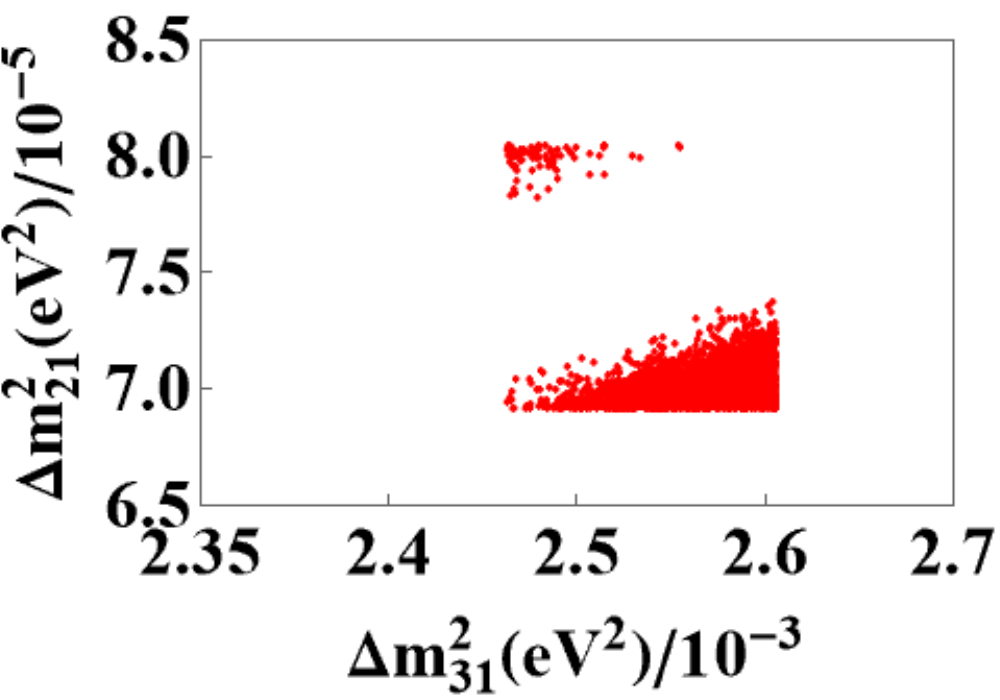}\label{fig:1(e)}}
    \subfigure[]{\includegraphics[width=0.48\textwidth]{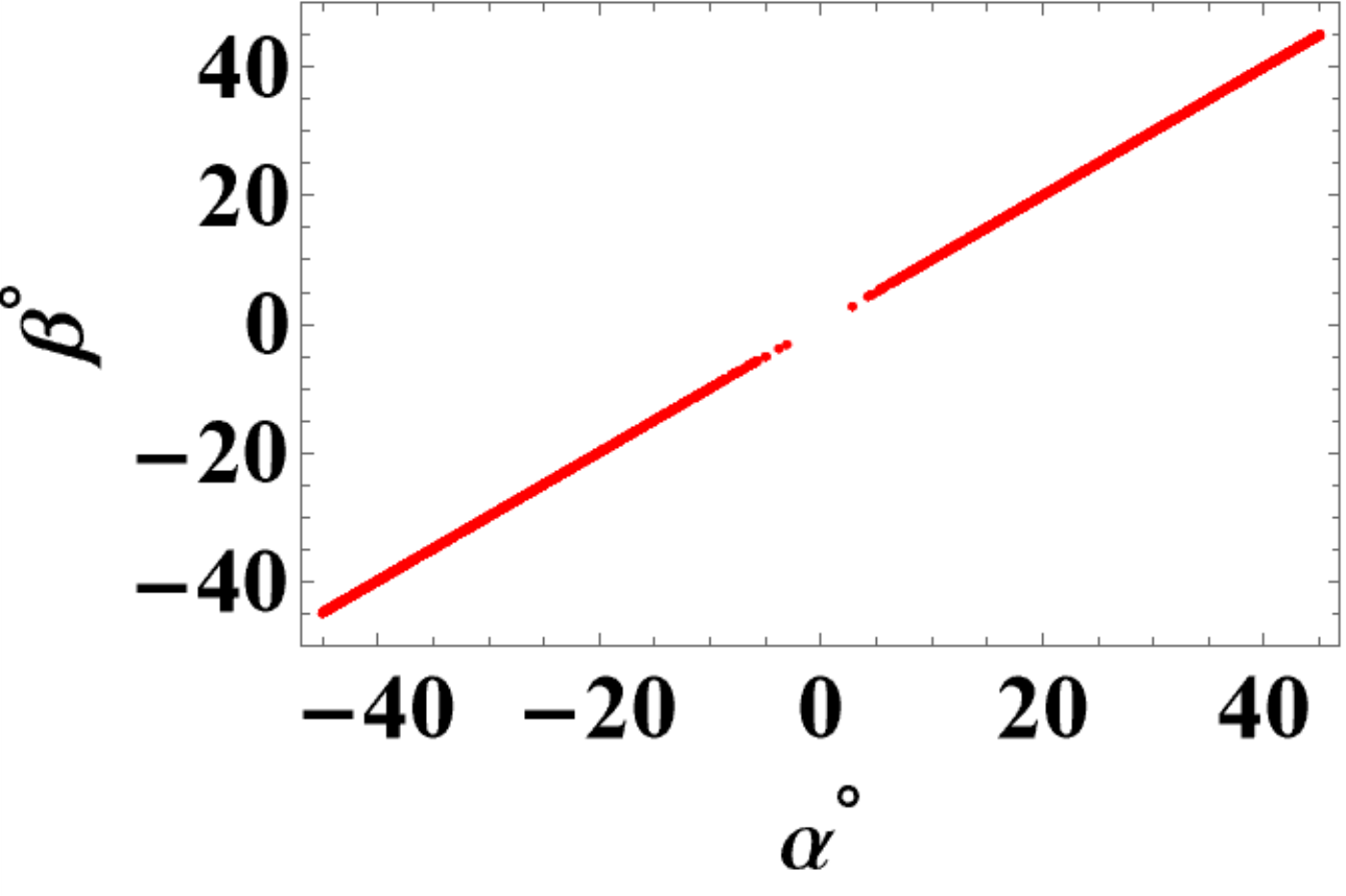}\label{fig:1(f)}}

    \caption{The correlation plots between (a) $\theta_{13}$ vs $\text{Re}[h]$ , (b) $\delta$ vs $\theta_{13}$, (c) $m_2$ vs $m_1$,  (d) $m_3$ vs $m_1$, (e) $\Delta m^2_{21}$ vs $\Delta m^2_{31}$ and (f) $\beta$ vs $\alpha$ for $\text{Re}[h]>0$ under the partial TBM scheme. }
\label{fig:1}
\end{figure}

\begin{figure}
  \centering
  \subfigure[]{\includegraphics[width=0.48\textwidth]{./hl0_thetavsh_FM.pdf}\label{fig:2(a)}}
  \subfigure[]{\includegraphics[width=0.48\textwidth]{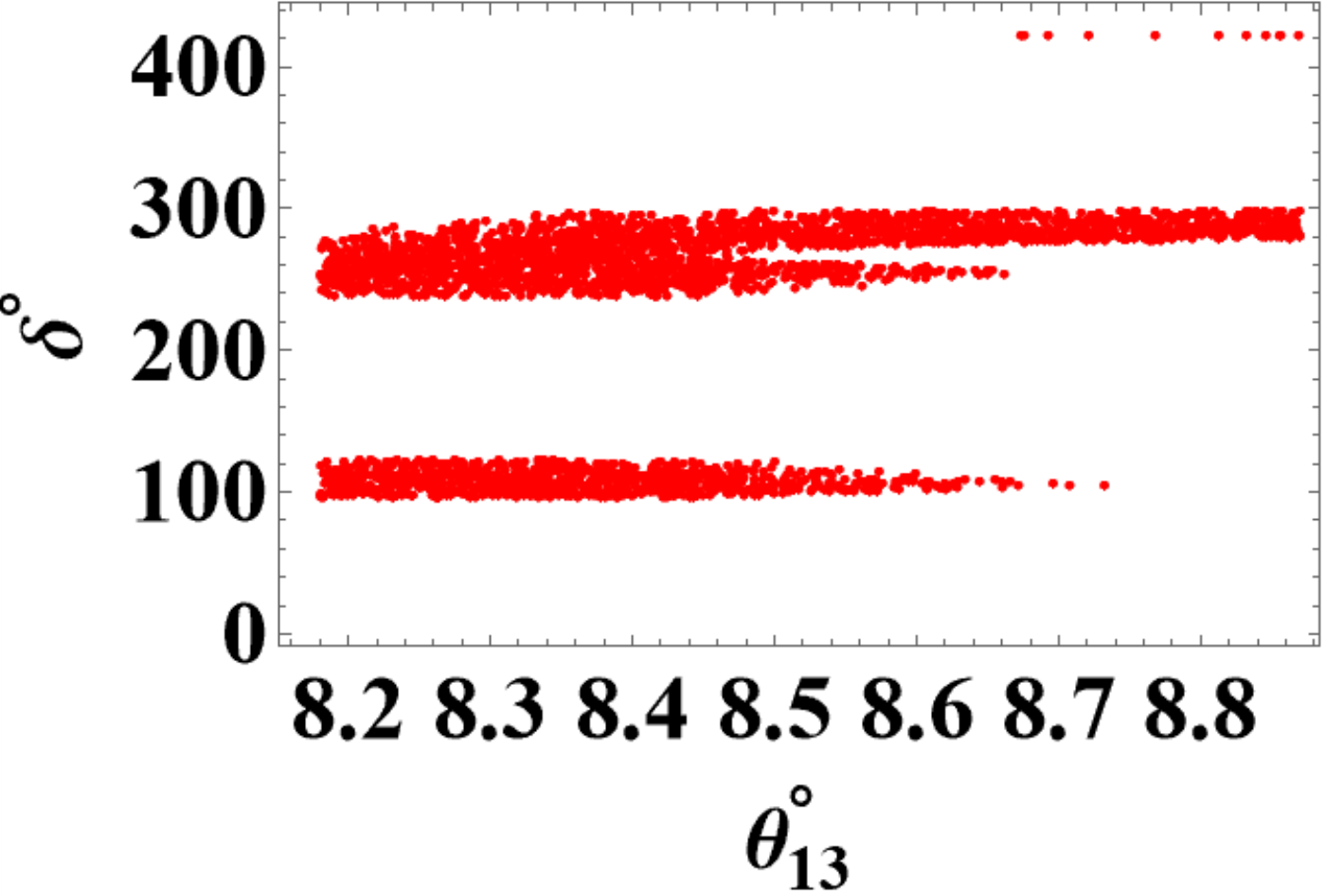}\label{fig:2(b)}}
   \subfigure[]{\includegraphics[width=0.48\textwidth]{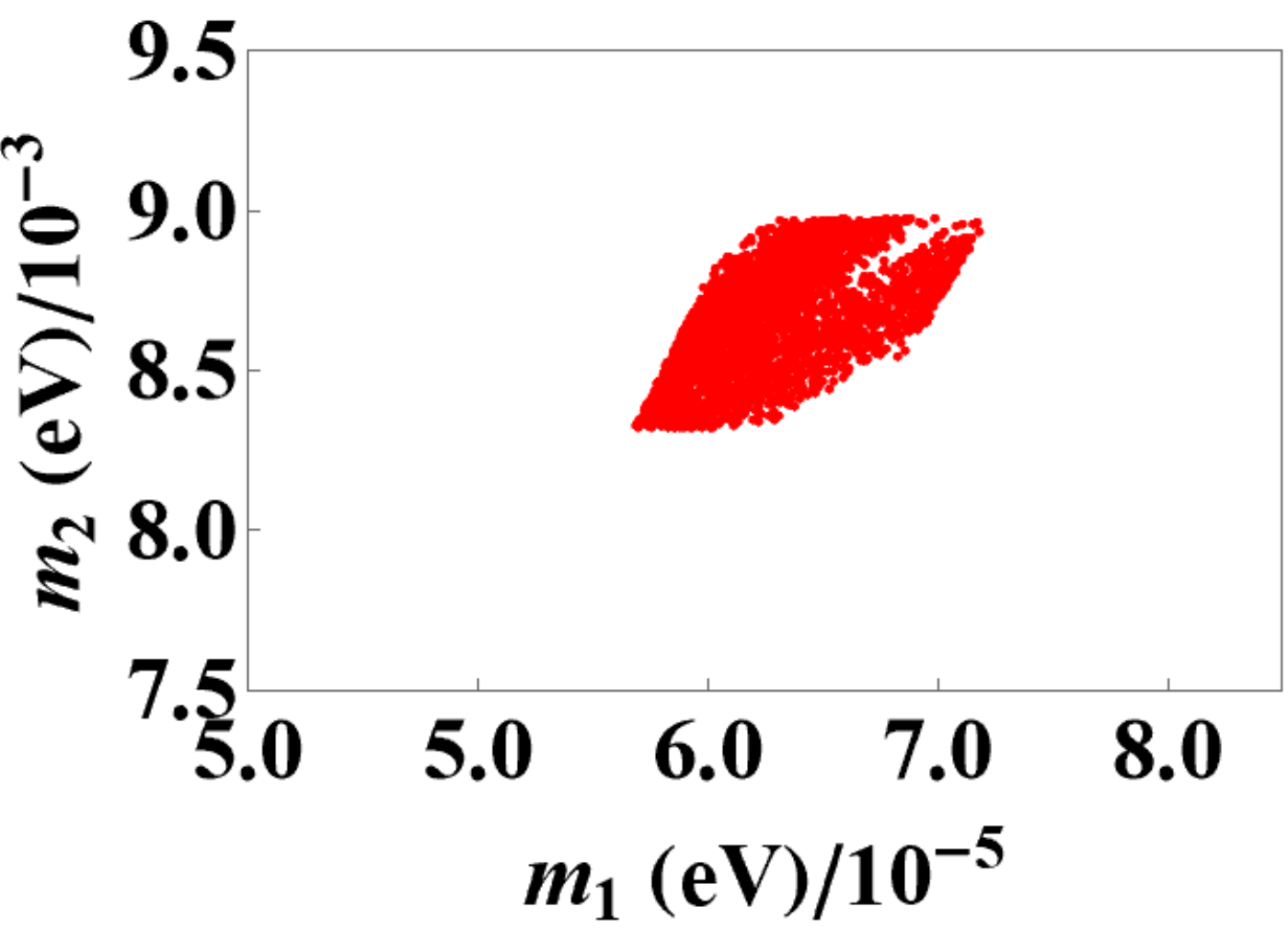}\label{fig:2(c)}} 
    \subfigure[]{\includegraphics[width=0.48\textwidth]{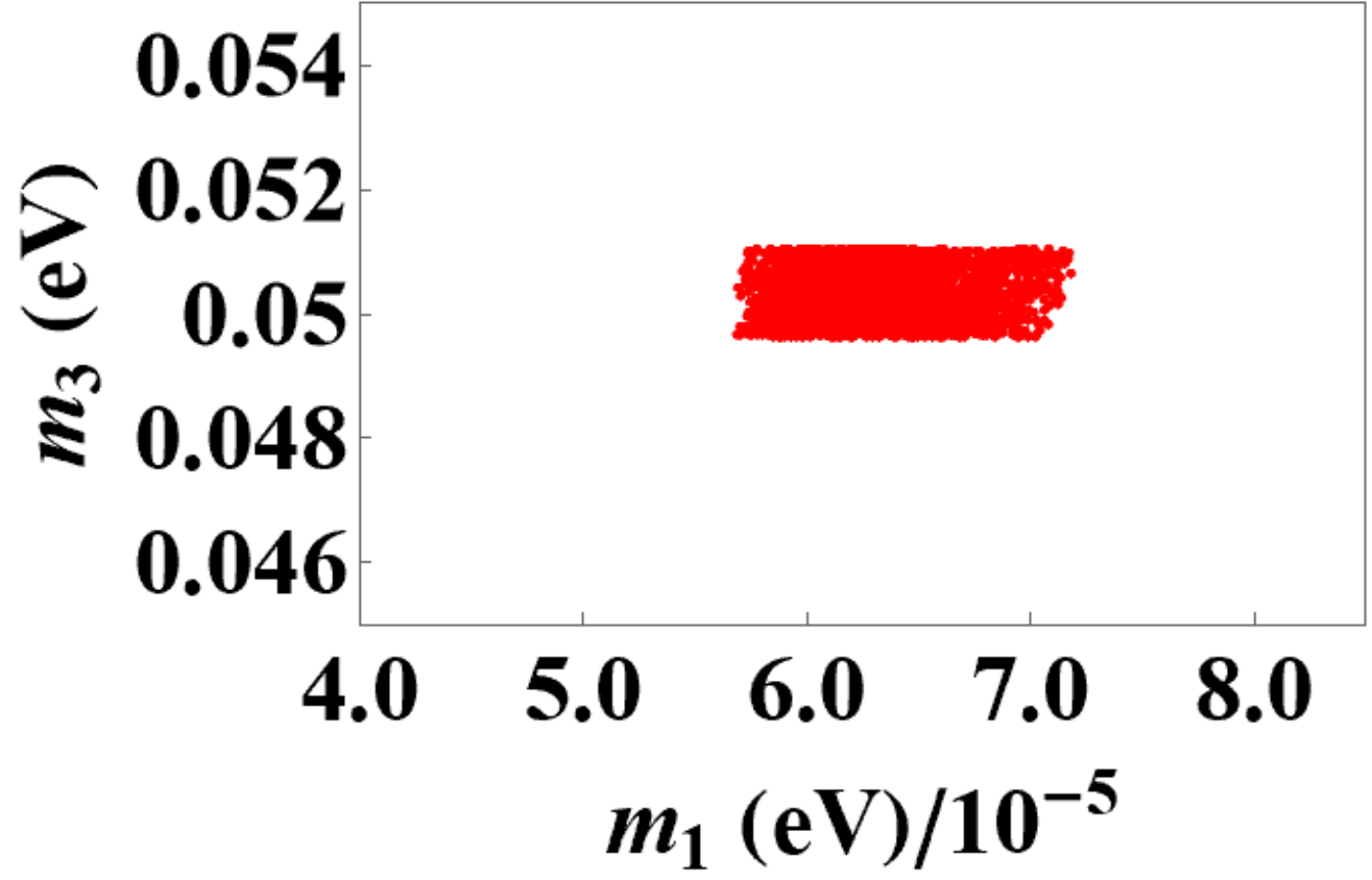}\label{fig:2(d)}}
     \subfigure[]{\includegraphics[width=0.48\textwidth]{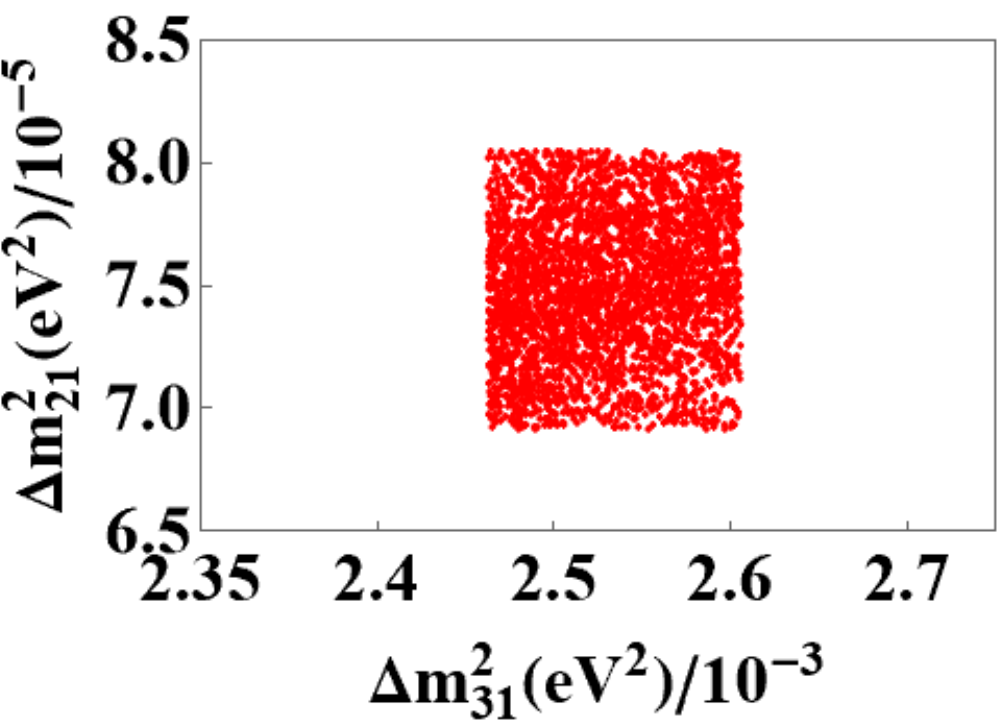}\label{fig:2(e)}}
    \subfigure[]{\includegraphics[width=0.48\textwidth]{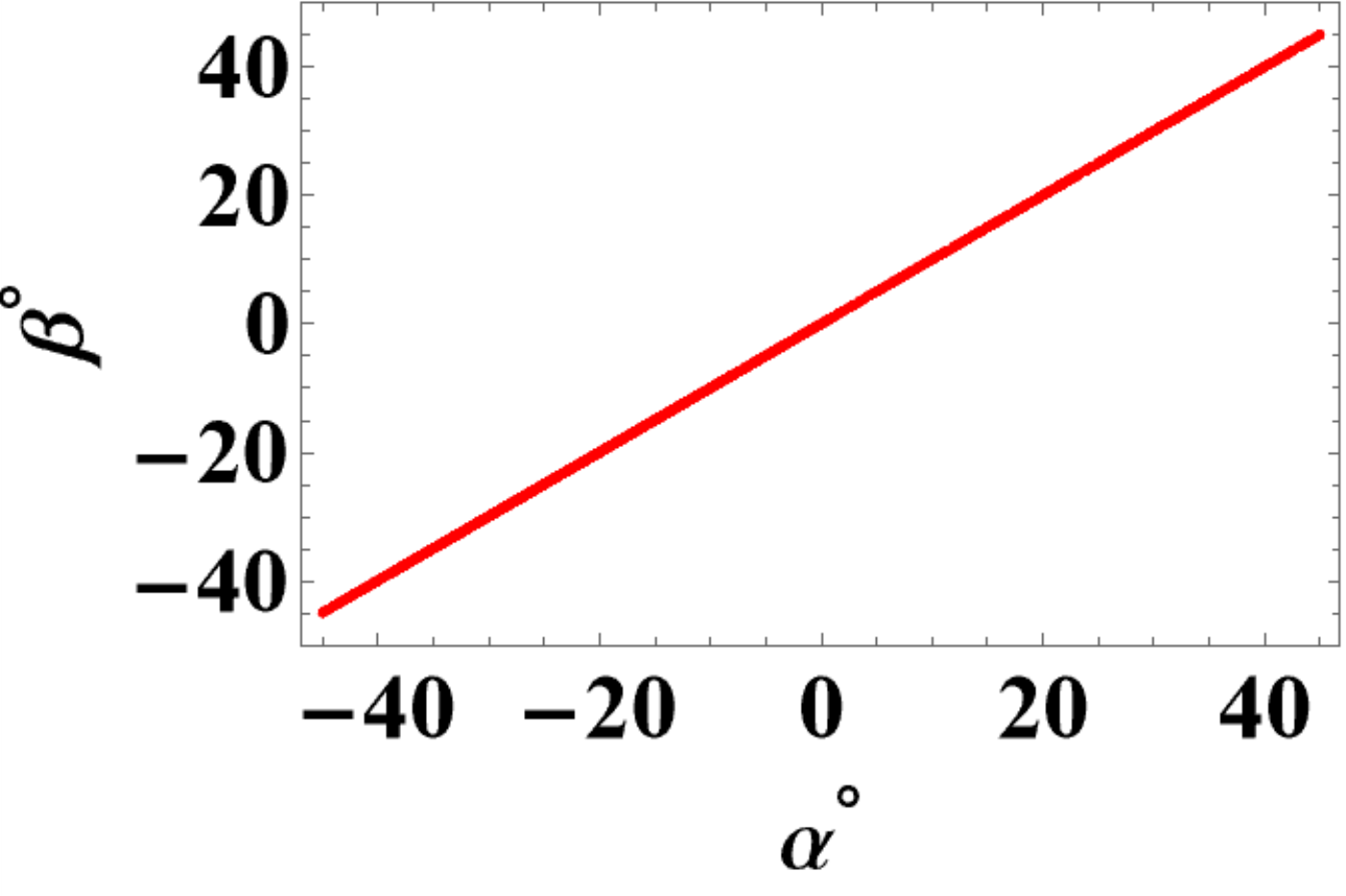}\label{fig:2(f)}}

    \caption{The correlation plots between (a) $\theta_{13}$ vs $\text{Re}[h]$ , (b) $\delta$ vs $\theta_{13}$, (c) $m_2$ vs $m_1$,  (d) $m_3$ vs $m_1$, (e) $\Delta m^2_{21}$ vs $\Delta m^2_{31}$ and (f) $\beta$ vs $\alpha$ for $\text{Re}[h]<0$ under the partial TBM scheme. }
\label{fig:2}
\end{figure}

First, we discuss the case, $\text{Re}[h]>0$ and find that many observable show peculiarities within their ranges. We summarise the main results as follows,
\begin{itemize}
\item We see that the result constrains $\text{Re}[h]$ to an even smaller interval of about $(1 \times10^{-4},6 \times 10^{-4})$, as shown in Fig.\,\ref{fig:1(a)}.
\item We see that the angle, $\theta_{13}$ shows a gap within its $3\sigma$ range\,\cite{Esteban:2024eli}, as can be seen from Figs. \ref{fig:1(a)} and \ref{fig:1(b)}. This gap is approximately from $8.26^\circ$ to $8.58^\circ$.
\item We can also see the same in the $3\sigma$ bounds\,\cite{Esteban:2024eli} of $\delta$. In fact, in this case, there are two such gaps, from $157^\circ$ to $194^\circ$ and $219^\circ$ to $316^\circ$, as seen from Fig.\,\ref{fig:1(b)}.
\item We see bounds on the values of $m_1$,\,$m_2$ and $m_3$ as $6.07 \times 10^{-5}\, \text{eV}<m_1<6.92 \times 10^{-5}\, \text{eV}$,\,$\,0.00827\, \text{eV}<m_2<0.00898\, \text{eV}$ and $0.049\, \text{eV}<m_3<0.051\, \text{eV}$ approximately(see Figs.\,\ref{fig:1(c)} and \ref{fig:1(d)}). Interestingly, $m_1$ and $m_2$ show some gaps in their values.  For $m_1$, the forbidden interval is $(6.28 \times 10^{-5}\, \text{eV}-6.32 \times 10^{-5}\, \text{eV})$, while for $m_2$, it is $(0.00863\, \text{eV}-0.00886\, \text{eV})$. 
\item We can see a similar pattern in the values of $\Delta m^2_{21}$, with the forbidden region lying from $7.45 \times 10^{-5} \,\text{eV}^2-7.83\times 10^{-5}\,\text{eV}^2$, as seen from Fig.\,\ref{fig:1(e)}.
\item Regarding the two Majorana phases, we observe that they are equal and span a range from about $-45^\circ$ to $45^\circ$. Further, they too exhibit an identical gap in their allowed ranges, approximately between \(0^\circ\) and \(5^\circ\) (see Fig.\,\ref{fig:1(f)}).
\end{itemize}

\begin{figure}
  \centering
    \subfigure[]{\includegraphics[width=0.48\textwidth]{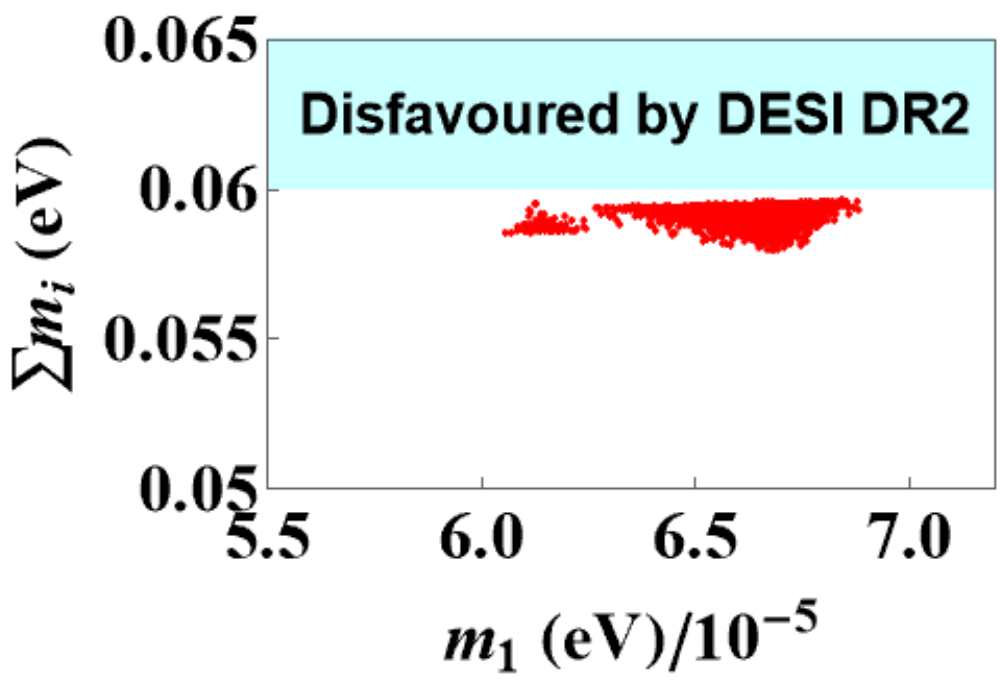}\label{fig:4(a)}} 
    \subfigure[]{\includegraphics[width=0.48\textwidth]{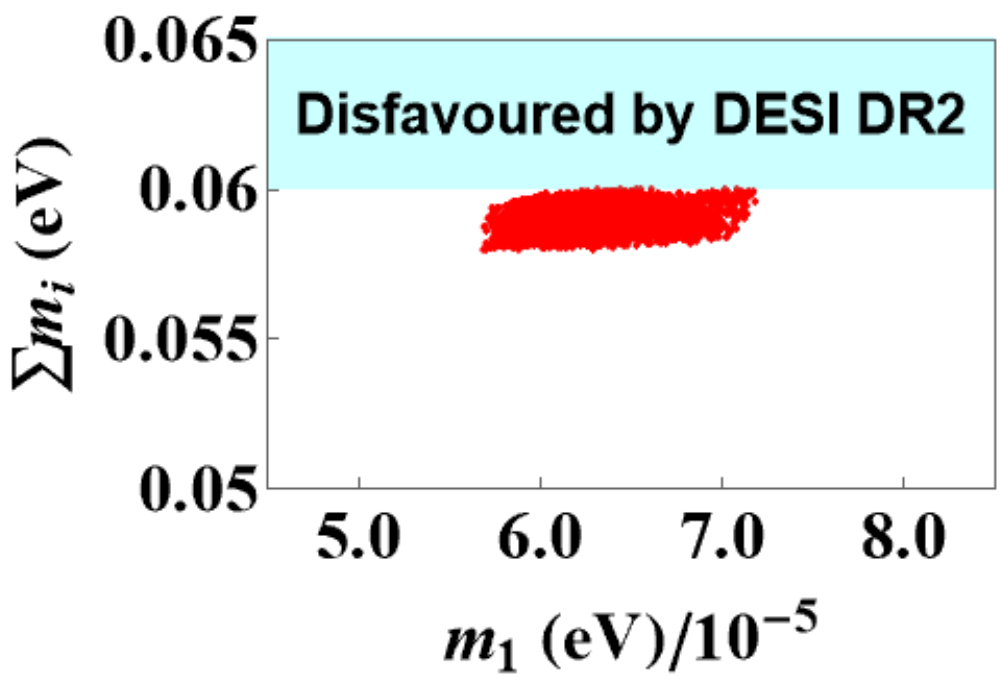}\label{fig:4(b)}} 
    \caption{The correlation plots between (a) $\Sigma\,m_i$ vs $m_1$ for $\text{Re}[h]>0$, (b) $\Sigma\,m_i$ vs $m_1$ for $\text{Re}[h]<0$ under the partial TBM scheme. }
\label{fig:4}
\end{figure}

\subsection*{Case II}

We now move to discuss the second case,  viz., $\text{Re}[h]<0$ and the main observations are presented here,
\begin{itemize}
\item Similar to Case 1, we see $\text{Re}[h]$ is constrained to a smaller interval of about $(-6 \times10^{-4},-3 \times 10^{-4})$, as shown in Fig.\,\ref{fig:2(a)}.
\item From Fig.\,\ref{fig:2(b)}, we observe that, unlike the previous case, the parameters here do not exhibit gaps in their allowed ranges, except for \(\delta\). The two forbidden regions for \(\delta\) lie approximately from \(122^\circ\) to \(233^\circ\) and $300^\circ$ to $420^\circ$ within the $3\sigma$ bounds\,\cite{Esteban:2024eli}.
\item The values of $m_1$,\,$m_2$ and $m_3$ are constrained to the approximate ranges $(5.74 \times 10^{-5}\,\text{eV}-7.26 \times 10^{-5}\,\text{eV}),\,(0.00830\,\text{eV}-0.00894\,\text{eV})$ and $(0.0495\,\text{eV}-0.0510\,\text{eV})$ respectively,  but without any forbidden regions, as observed in Figs. \,\ref{fig:2(c)} and \ref{fig:2(d)}.
\item We can see that $\Delta m^2_{21}$ and $\Delta m^2_{21}$ are consistent with experimental data\,(see Fig.\,\ref{fig:2(e)}). 
\item Like in the previous case, the two Majorana phases are equal and have the same ranges but without any forbidden gaps in their ranges (see Fig.\,\ref{fig:2(f)}).
\end{itemize}

In the present work, the sum of three neutrino mass $\sum m_i$ aligns with Plank's data\,($<0.12$\,eV)\,\cite{Planck:2018vyg}. Recently, the DESI-DR2 has updated the bound on $\sum m_i$ which is $<0.06$ eV\,\cite{DESI:2025zgx}. Our analysis is also consistent with the updated cosmological constraint. The graphical visualization is shown in Fig.\,(\ref{fig:4}).

The neutrino-less double beta decay ($0\nu\beta\beta$)\cite{Schechter:1981bd}, is a lepton number violating decay and its discovery will justify the Majorana nature of neutrinos. The decay rate $\Gamma$ of the said process is determined by the phase space factor $G^{0\nu}$, the nuclear matrix element $M^{0\nu}$, and the effective Majorana mass $m_{\beta\beta}$, following the relation: $\Gamma\sim G^{0\nu}. M^{0\nu}.m^2_{\beta\beta}$.
 
\begin{figure}
  \centering
    \subfigure[]{\includegraphics[width=0.48\textwidth]{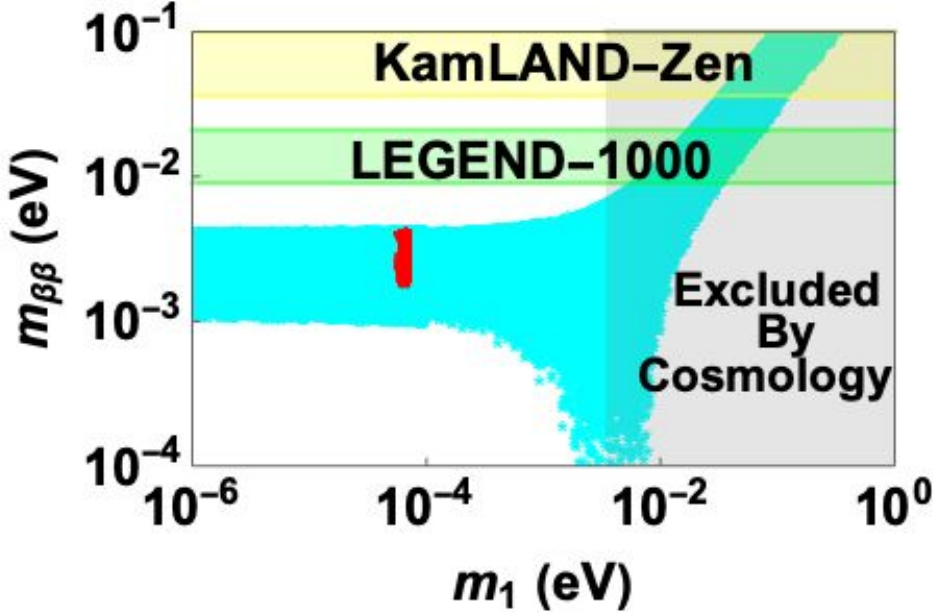}\label{fig:4(a)}} 
    \subfigure[]{\includegraphics[width=0.48\textwidth]{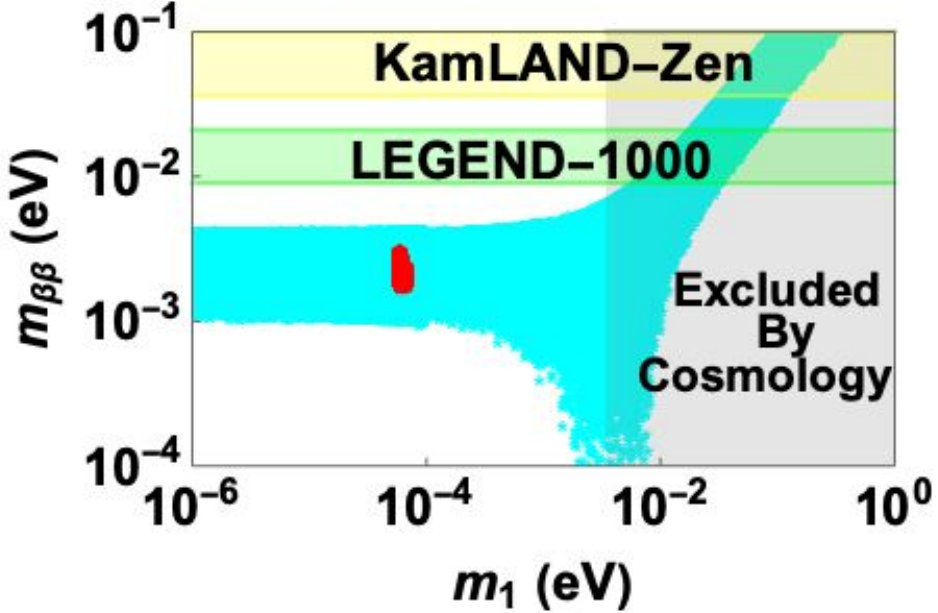}\label{fig:4(b)}} 
    \caption{The correlation plots between (a) $m_{\beta\beta}$ vs $m_1$ for $\text{Re}[h]>0$, (b) $m_{\beta\beta}$ vs $m_1$ for $\text{Re}[h]<0$ under the partial TBM scheme. }
\label{fig:5}
\end{figure}

The effective Majorana neutrino mass $m_{\beta \beta}$ is an observational parameter which is expressed as: \[m_{\beta \beta}=|\sum_{k=1}^{3}{U^2_{1k}m_k}|.\] where, $m_1$, $m_2$ and $m_3$ are the three mass eigenvalues. The $U_{11}$, $U_{12}$ and $U_{13}$ are the elements of the PMNS matrix that contains the information of the Majorana phases. Several experiments have provided the upper bounds of $m_{\beta \beta}$: SuperNEMO(Se$^{82}$) as $67-131$ meV,  GERDA(Ge$^{76}$) as $104-228$ meV, EXO-200(Xe$^{136}$) as $111-477$ meV, CUORE(Te$^{130}$) as $75-350$ meV and KamLAND-Zen(Xe$^{136}$) as $61-165$ meV \cite{Ejiri:2020xmm, Agostini:2022zub, CUORE:2019yfd, GERDA:2019ivs, KamLAND-Zen:2016pfg, SuperNEMO:2021hqx, CUORE:2018ncg}. In this regard, we visualize the prediction of the parameter $m_{\beta \beta}$ from the predictions of our studied texture\,(see Fig.\,\ref{fig:5}). We see that that the prediction of $m_{\beta \beta}$ might fall within the sensitivity of future experiments for both cases.

 In the next section, we outline a symmetry framework that offers a possible origin for the proposed texture.

\section{Symmetry Perspective \label{sec5}}

In this section, we explore how the proposed texture (Eq.\,(\ref{Mnu})) may be realized from first principles via a combination of one Type-I and two Type-II seesaw mechanisms\,\cite{Chakraborty:2024eki}. For this, we extend the SM scalar sector by seven scalars and the group structure by a non-Abelian discrete symmetry group, $A_4$ and two cyclic groups, $Z_7$ and $Z_{10}$.  The charge assignments of the fields under the extended group $SU(2)_L \otimes U(1)_Y \otimes A_4 \otimes Z_{10} \otimes Z_{7}$ are given in Table\,\ref{Field Content of M}.
The Yukawa Lagrangian under the said group is constructed in the following way,

\begin{align}
- \mathcal{L}_Y =& y_{e}(\bar{D}_{l_{L}}H) e_{R} + y_{\mu}(\bar{D}_{l_{L}}H)\mu_{R} + y_{\tau}(\bar{D}_{l_{L}}H)\tau_{R}+ \,y_{1}(\bar{D}_{l_{L}}\tilde{H})\nu_{eR}+
\nonumber \\
&\frac{y_2}{\Lambda}(\bar{D}_{l_{L}}\tilde{H})\nu_{\mu_R} \chi + \,\frac{y_3}{\Lambda} (\bar{D}_{l_{L}}\tilde{H})\nu_{\tau_R}\kappa
+ \frac{y_{c}}{2}(\overline{\nu^c_{e_{R}}}
\nu_{e_{R}})\eta +\frac{y_{a}}{2}(\overline{\nu^c_{\mu_{R}}}\nu_{\mu_{R}})\nonumber \\
&\xi
+ \frac{y_{b}}{2} [(\overline{\nu^c_{e_{R}}}\nu_{\tau_{R}})+ (\overline{\nu^c_{\tau_{R}}}\nu_{e_{R}})]\xi+ \frac{y_{d}}{2}(\overline{\nu^c_{\tau_{R}}}\nu_{\tau_{R}})\psi + y_{t}(\bar{D}_{l_{L}} D_{l_{L}}^{c})i \sigma_2 \tilde{\Delta}_1 \nonumber \\
&
 +\frac{y_{s}}{\Lambda}(\bar{D}_{l_{L}} 
D_{l_{L}}^{c})\,i \sigma_2 \tilde{\Delta}_2 \xi + h.c.,
\label{ Lagrangian}
\end{align}

where $\Lambda$ is the cut-off scale of the theory. The auxiliary groups $Z_7$ and $Z_{10}$ are introduced to eliminate certain undesired terms that are allowed by $A_4$. The group $Z_7$ is instrumental in forbidding Weinberg's dimension 5 operator\,$\frac{1}{\Lambda}\bar{D}_{l_{L}}\tilde{H}\\\tilde{H}^TD_{l_{L}}^c$ and other similar terms\,\cite{Vien:2025fiu, Chakraborty:2024hhq, Chakraborty:2025juy}. The product rules under the $A_4$ group are given in Appendix \ref{appendix a}. 

\begin{table}
\centering
\begin{tabular*}{\textwidth}{@{\extracolsep{\fill}} cccccccccccc}
\hline\hline
Fields & $D_{l_{L}}$ &  $(e_R, \mu_R, \tau_R)$ & $(\nu_{e_R}, \nu_{\mu_R}, \nu_{\tau_R})$& $H$  &$\Delta_1$&$\Delta_2$& $\xi$ & $\eta$ &  $\chi$ & $\psi$ & $\kappa$ \\ 
\hline\hline
$SU(2)_L$ & 2 & $(1,1,1)$ & $(1,1,1)$& 2 & 3 & 3 & 1 & 1 & 1 & 1 & 1  \\
\hline
$U(1)_Y$ & -1 & $(-2,-2,-2)$ & $(0,0,0)$ & 1 & -2 & 2 & 0 & 0 & 0 & 0 & 0 \\
\hline
$A_4$ & 3 &$ (1, 1^{''} , 1^{'})$ & $(1^{''}, 1 , 1^{'})$ & 3 & 3 & 3 & 1 & $1^{''}$ & 1 & $1^{'}$ & 1 \\
\hline 
$Z_{10}$ & 0 & $(0,0,0)$ & $(0,4,8)$ & 0 & 0 & 8 & 2 & 0 & 6 & 4 & 2 \\
\hline
$Z_7$ & -2 & $(4,4,4)$ & $(6,6,6)$ & 1 & 3 & 1 & 2 & 2 & 0 & 2 & 0 \\
\hline
\end{tabular*}
\caption{ The transformation properties of various fields under $SU(2)_L \otimes U(1)_Y \otimes A_{4} \otimes Z_{10} \otimes Z_{7}$.}
\label{Field Content of M}
\end{table} 

To obtain the proposed texture, the vacuum expectation values\,(vev) for the scalar fields are chosen as: $\langle H \rangle_{0}=(1,0,0)^T v_{H}$, \,$\langle \Delta_1 \rangle_{0}=(1,-1,-1)^T v_{1}$,\,$\langle \Delta_2 \rangle_{0}=(0,1,-1)^T v_{2}$,\,$\langle \xi \rangle_{0}=v_{\xi}$,\,$\langle \eta \rangle_{0}=v_{\eta}$,\,$\langle \chi \rangle_{0}=v_{\chi}$,\,$\langle \psi \rangle_{0}=v_{\psi}$ and $\langle \kappa \rangle_{0}=v_{\kappa}$. 

The charged lepton mass matrix can be derived from Eq.(\ref{ Lagrangian}) as shown in the following,

\begin{equation}
 M_{l} = 
 \begin{bmatrix}
 y_e \,v_{H} &  0 &  0\\
 0 & y_\mu \,v_{H} & 0\\
 0 & 0 & y_\tau \,v_{H}\\
 \end{bmatrix},\nonumber
 \label{chargrd lepton mass matrix}
\end{equation}

The Dirac neutrino mass matrix ($M_D$) and right handed neutrino mass matrix ($M_R$) as shown below,

\begin{eqnarray}
 M_{D} &=&
 \begin{bmatrix}
  0 &  \frac{y_2}{\Lambda}\,v_H v_\chi & 0\\
 y_1\,v_H & 0 & 0 \\
 0 & 0  & \frac{y_3}{\Lambda}\,v_H \,v_\kappa\\
 \end{bmatrix},
 \label{Dirac Mass Matric M1}
 \quad M_{R}=
\begin{bmatrix}
 \frac{y_c}{2} v_{\eta } & 0 & \frac{y_b}{2} v_{\xi } \\
 0 & \frac{y_a}{2} v_{\xi } & 0 \\
\frac{y_b}{2} v_{\xi } & 0 & \frac{y_d }{2} v_{\psi } \\
\end{bmatrix},\nonumber
 \label{M_R for M1}
\end{eqnarray}

The two matrices $M^1_{T_2}$ and $M^2_{T_2}$ from two Type-II seesaw mechanisms can be derived as shown in the following

\begin{eqnarray}
 M^1_{T_2} &=&
 \begin{bmatrix}
 \frac{2 }{3} v_1 y_t & \frac{1}{3} v_1 y_t & \frac{1}{3} v_1 y_t \\
 \frac{1}{3} v_1 y_t & -\frac{2}{3} v_1 y_t & -\frac{1}{3} v_1 y_t \\
 \frac{1}{3} v_1 y_t & -\frac{1}{3} v_1 y_t & -\frac{2}{3} v_1 y_t \\
 \end{bmatrix},
 \label{first type-2}\quad
 M^2_{T_2}=
 \begin{bmatrix}
 0 & \frac{1}{3 \Lambda } v_2 y_s v_{\xi } & -\frac{1}{3 \Lambda } v_2 y_s v_{\xi }\\
 \frac{1}{3 \Lambda } v_2 y_s v_{\xi } & \frac{2 }{3 \Lambda } v_2 y_s v_{\xi }& 0 \\
 -\frac{1}{3 \Lambda } v_2 y_s v_{\xi } & 0 & -\frac{1}{3 \Lambda } 2 v_2 y_s v_{\xi } \\
 \end{bmatrix}.\nonumber
 \label{second type-2}
\end{eqnarray}

 The effective neutrino mass matrix $M$ is constructed after taking the contributions from Type-I ($-\,M_{D}M^{-1}_{R}M^{T}_{D}$) and two Type-II seesaw mechanisms involving the vevs of the scalar fields. The $M$ becomes a function of the parameters:
$y_1,y_2,y_3,y_a,y_b,y_c,y_d,y_t, 
y_s,\Lambda,v_H,
v_1,v_2,v_\chi,v_\xi,v_\kappa,v_\psi,v_\eta$ and
will reduce to the proposed texture $M_\nu$ (Eq.\,(\ref{Mnu})) with the following constraints,
\begin{eqnarray}
D_1 &=& y_t v_1 = 3b,\nonumber \\
D_2 &=& \frac{y_s v_2 v_\xi}{\Lambda} = 3a,\nonumber \\
D_3 &=& \frac{y_d \Lambda^2 v_\psi}{v_H^2 v^2_\kappa y^2_3} = \frac{2(2b-a+g+h)}{[4a^2-6ab-3b^2-2bg+h^2-4a(g+h)]},\nonumber \\
D_4 &=& \frac{y_a \Lambda^2 v_\xi}{v^2_H v^2_\chi y^2_2} = \frac{10}{5a-10b+9h},\nonumber \\
D_5 &=& \frac{y_b \Lambda v_\xi}{v^2_H v_\kappa y_1 y_3} = \frac{2(a-b-g)}{[4a^2-6ab-3b^2-2bg+h^2-4a(g+h)]}, \nonumber \\
D_6 &=& \frac{y_c v_\psi}{v^2_H v^2_\kappa y^2_3} = \frac{2(3a+2b+g-h)}{[4a^2-6ab-3b^2-2bg+h^2-4a(g+h)]}.\nonumber
\end{eqnarray}

The $SU(2)_L \otimes U(1)_Y \otimes A_{4} \otimes Z_{10} \otimes Z_{7}$ invariant scalar potential, along with the optimization conditions for the given choice of vevs, is discussed in Appendix \ref{appendix b}. It is important to mention that the inclusion of the Higgs triplet scalar in the theory leads to a new term in the scalar potential $(H, \Delta)$ including interactions associated with $\Delta$ that violate the global lepton number symmetry. The linear term\,$(\mu H^T i \sigma_2 \Delta^\dagger H)$ is crucial to obtain the small vev of scalar triplets responsible for the Type-II seesaw mechanism. The cubic $\mu$ dependent term is important for determining the VEV obtained by the neutral component\,($\Delta^0$) of the scalar triplet. In our model, the said term does not appear at the tree level due to our choice of charges under the several cyclic groups. However, the contributions come from the next to leading order terms $Y_1 \eta ( H^T i \sigma_2 \Delta_1^\dagger H )$ and $\frac{Y_2}{\Lambda} \psi \eta ( H^T i \sigma_2 \Delta_2^\dagger H)$ respectively\,(see Appendix \ref{appendix b}). Needless to mention, the cubic $\mu$ for the terms are replaced by $Y_1 \eta$ and $\frac{Y_2}{\Lambda} \psi \eta$ respectively.

We now proceed to summarize the key findings of our work.

\section{Summary and Discussion \label{sec6}}
 A new, minimal and predictive Majorana neutrino mass matrix texture is posited. Its various predictions are studied under a new mixing scheme: a partial TBM mixing where $\sin\,\theta_{12}=1/\sqrt{3}$ and $\sin\,\theta_{23}=1/\sqrt{2}$, but $\theta_{13}$ and $\delta$ remain as free parameters. The texture rules out any possibility of $\theta_{13}$ being zero and it is not a $\mu-\tau$ deviated texture. As an interesting feature, the proposed texture predicts equal Majorana phases. We have seen that the observable parameters predicted by the texture depend on three real parameters, viz.,$\text{Re}[h],\,\theta_{13}$ and $\delta$. We have randomized $\theta_{13}$ and $\delta$ within their $3\sigma$ bounds and found two interesting cases emerge depending on the values of $\text{Re}[h]$. We have observed that when $\text{Re}[h]>0$, some forbidden zones appear in the ranges for many observable parameters such as $\theta_{13},\,\delta,\,m_1,\,m_2,\,\Delta m^2_{21}, \,\alpha$ and $\beta$. But such gaps drastically disappear when we consider the case, $\text{Re}[h]<0$, with only $\delta$ showing one. It is important to emphasize that the sign of $\text{Re}[h]$ is not associated with distinct symmetry realizations of the model, but instead reflects a choice within the parameter space of our analysis. Nevertheless, the resulting phenomenology exhibits a non-trivial dependence on the sign of $\text{Re}[h]$ . In particular, we find that $\text{Re}[h] > 0$ leads to more restrictive bounds on the physical observables compared to the case $\text{Re}[h]<0$.
 
The texture may be realized from first principle with some reasonable constraints on model parameters.  We have shown it under an extended symmetry group $SU(2)_L \otimes U(1)_Y \otimes A_4 \otimes Z_{10} \otimes Z_{7}$ based on a hybrid framework of one Type-I seesaw and two Type-II seesaw mechanisms. 

It is important to mention that the recent results from JUNO experiment have significantly improved the precision of the solar mixing angle $\theta_{12}$\,\cite{Capozzi:2025ovi, JUNO:2025gmd, Esteban:2026phq}. In particular, the strict tri-bimaximal prediction of $\sin \theta_{12} = 1/\sqrt{3}$ is now disfavoured\,\cite{Capozzi:2025ovi, JUNO:2025gmd, Esteban:2026phq}. This can be understood as a deviation from the tri-bimaximal value\,($\sin^2 \theta_{12} = 1/3 + \epsilon$), with $\epsilon \approx -0.025$. Such a small deviation, consistent with the NuFIT~6.1 data, allows the proposed texture to remain compatible with current experimental observations, although the allowed parameter space becomes more constrained. However, the changes are very small, proportional to $\mathcal{O}(\epsilon)$ or $\mathcal{O}(\epsilon^2)$. 
 
 The present model shelters several scalar fields, which play a crucial role in maintaining the independence of the texture parameters while expressing them in terms of the model parameters. In addition, to cut short certain terms from the right-handed neutrino mass matrix $(M_R)$, we introduce some extra scalar fields in our model. This step is necessary to enhance the efficiency of the model. In this regard, the inclusion of the scalar fields becomes inevitable. A more minimal model might be achieved with fewer number of scalar fields with Weinberg like terms instead of Type-II seesaw contributions. However, this is a part of our future prospects and beyond the scope of the present work. 
 
The texture's phenomenological predictions make it a compelling framework for further investigation, particularly in the context of leptogenesis and its connection to the observed baryon asymmetry of the Universe.

\section{Acknowledgement}

STG thanks Nihar Ranjan Saha, Indian Institute of Technology, Madras for a fruitful discussion on some aspects of numerical analysis. 

\section{Funding Information}

The research work of Pralay Chakraborty is supported by Innovation in Science Pursuit for Inspired Research (INSPIRE), Department of Science and Technology, Government of India, New Delhi vide grant No. IF190651.

\appendix

\section{Product Rules of $A_4$ \label{appendix a}}

The non-Abelian discrete group, $A_4$ has four inequivalent irreducible representations,viz., three singlets, $1,\,1^{'}$ and $1^{''}$ and one triplet, $3$.  In the Altarelli-Feruglio basis\,\cite{Altarelli:2005yx,Ishimori:2010au}, the multiplication rules take the following form:
\begin{align}
\begin{pmatrix}
a_1\\
a_2\\
a_3
\end{pmatrix}_3 \otimes \begin{pmatrix}
b_1\\
b_2\\
b_3
\end{pmatrix}_3 =& (a_1 b_1 +a_2 b_3+a_3 b_2)_1\oplus (a_3 b_3+a_1 b_2+a_2 b_1)_{1^{'}} \oplus (a_2 b_2+a_1 b_3+\nonumber \\&a_3 b_1)_{1^{''}} 
 \oplus \frac{1}{3} \begin{pmatrix}
2a_1 b_1-a_2 b_3-a_3 b_2\\
2a_3 b_3-a_1 b_2-a_2 b_1\\
2a_ 2b_2-a_1 b_3-a_3 b_1\\
\end{pmatrix}_{3_{s}} \oplus \frac{1}{2}\begin{pmatrix}
a_2 b_3-a_3 b_2\\
a_1 b_2-a_2 b_1\\
a_1 b_3-a_3 b_1
\end{pmatrix}_{3_{a}},\nonumber
\end{align}
and, 
\begin{align}
1 \otimes 1=&1,\quad
1^{'} \otimes 1^{''}= 1,\quad
1^{''} \otimes 1^{'}=1. \nonumber
\end{align}

\section{The Scalar Potential \label{appendix b}}

To justify the vacuum alignments of the scalar fields, we construct the $SU(2)_L \otimes U(1)_Y \otimes A_4  \otimes Z_{10} \otimes Z_{7}$ invariant scalar potential as shown below,

\begin{align}
V =& V(H)+V(\psi)+V(\eta)+V(\chi)+V(\kappa)+V(\xi)+V(\Delta_1)+V(\Delta_2)\nonumber \\
&+V(H,\psi)+\,V(H, \eta)+V(H, \chi)
+\,V(H, \kappa) + V(H, \xi)+V(H,\Delta_1)\nonumber \\
&+V(H,\Delta_2)+V(\psi, \eta)+V(\psi, \chi)+V(\psi, \kappa)+V(\psi, \xi)+V(\psi, \Delta_1)\nonumber \\
&+V(\psi, \Delta_2)+V(\eta, \chi)+V(\eta, \kappa)+V(\eta, \xi)+V(\eta, \Delta_1)+V(\eta, \Delta_2)+\nonumber \\
&V(\chi, \kappa)+V(\chi, \xi)+V(\chi, \Delta_1)
+V(\chi, \Delta_2)+V(\kappa, \xi)+V(\kappa, \Delta_1)
+\nonumber \\
&V(\kappa, \Delta_2)+V(\xi, \Delta_1)+V(\xi, \Delta_2)+V(\Delta_1, \Delta_2)+h.c.\nonumber
\end{align}

The explicit forms of the terms appearing in the scalar potential are shown in the following,

\begin{align}
V(H) =& -\mu_H^2 (H^\dagger H) 
+ \lambda_1^H (H^\dagger H)(H^\dagger H) 
+ \lambda_2^H (H^\dagger H)_{1'} (H^\dagger H)_{1''} 
+ \lambda_3^H (H^\dagger H)_{3s} (H^\dagger \nonumber \\
&H)_{3s} 
+ \lambda_4^H (H^\dagger H)_{3s} 
(H^\dagger H)_{3a} + \lambda_5^H (H^\dagger H)_{3a} (H^\dagger H)_{3a}
,\nonumber\\
V(\psi) =& -\mu_\psi^2 \, (\psi^\dagger \psi) 
+ \lambda^\psi \, (\psi^\dagger \psi)(\psi^\dagger \psi)
,\nonumber 
\quad
V(\eta) = -\mu_\eta^2 \, (\eta^\dagger \eta) 
+ \lambda^\eta \, (\eta^\dagger \eta)(\eta^\dagger \eta),\nonumber 
\\
V(\chi)=& -\mu^2_\chi (\chi^\dagger \chi)+\lambda^\chi (\chi^\dagger \chi)(\chi^\dagger \chi),\nonumber 
\quad
V(\kappa)= -\mu^2_\kappa (\kappa^\dagger \kappa)+\lambda^\kappa (\kappa^\dagger \kappa)(\kappa^\dagger \kappa),\nonumber \\
V(\xi)=& -\mu^2_\xi (\xi^\dagger \xi)+\lambda^\xi (\xi^\dagger \xi)(\xi^\dagger \xi),\nonumber 
\quad
V(\xi,\Delta_1)= \lambda^{ \xi \Delta_1}_1 Tr(\xi^\dagger \xi)(\Delta_1^\dagger \Delta_1),\nonumber \\
%
V(\Delta_1) =&- \mu_{\Delta_1}^2 \, \text{Tr}(\Delta_1^\dagger \Delta_1) 
+ \lambda_1^{\Delta_1} \, \text{Tr}(\Delta_1^\dagger \Delta_1)\, \text{Tr}(\Delta_1^\dagger \Delta_1) 
+ \lambda_2^{\Delta_1} \, \text{Tr}(\Delta_1^\dagger \Delta_1)_{1'}\, \text{Tr}(\Delta_1^\dagger \Delta_1)_{1''} 
\nonumber \\
&+ \lambda_3^{\Delta_1} \text{Tr}(\Delta_1^\dagger \Delta_1)_{3s}\, 
\text{Tr}(\Delta_1^\dagger \Delta_1)_{3s} + \lambda_4^{\Delta_1} \, \text{Tr}(\Delta_1^\dagger \Delta_1)_{3s}\, \text{Tr}(\Delta_1^\dagger \Delta_1)_{3a} 
+ \lambda_5^{\Delta_1} \, \text{Tr}(\Delta_1^\dagger \nonumber \\
&\Delta_1)_{3a}\, \text{Tr}(\Delta_1^\dagger \Delta_1)_{3a}
,\nonumber \\
V(\Delta_2) =&- \mu_{\Delta_2}^2 \, \text{Tr}(\Delta_2^\dagger \Delta_2) 
+ \lambda_1^{\Delta_2} \, \text{Tr}(\Delta_2^\dagger \Delta_2)\, \text{Tr}(\Delta_2^\dagger \Delta_2) 
+ \lambda_2^{\Delta_2} \, \text{Tr}(\Delta_2^\dagger \Delta_2)_{1'}\, \text{Tr}(\Delta_2^\dagger \Delta_2)_{1''} 
\nonumber \\
&+ \lambda_3^{\Delta_2} \, \text{Tr}(\Delta_2^\dagger \Delta_2)_{3s} 
\text{Tr}(\Delta_2^\dagger \Delta_2)_{3s} + \lambda_4^{\Delta_2} \, \text{Tr}(\Delta_2^\dagger \Delta_2)_{3s}\, \text{Tr}(\Delta_2^\dagger \Delta_2)_{3a} 
+ \lambda_5^{\Delta_2} \, \text{Tr}(\Delta_2^\dagger\nonumber \\
& \Delta_2)_{3a}\, \text{Tr}(\Delta_2^\dagger \Delta_2)_{3a}
,\nonumber \\
V(H, \psi)=& \lambda^{H \psi}_1 (H^\dagger H)(\psi^\dagger \psi),\nonumber
\,\,
V(H, \eta)= \lambda^{H \eta}_1 (H^\dagger H)(\eta^\dagger \eta),\nonumber
\,\,
V(H, \chi)= \lambda^{H \chi}_1 (H^\dagger H)(\chi^\dagger \chi),\nonumber \\
V(H, \kappa)=& \lambda^{H \kappa}_1 (H^\dagger H)(\kappa^\dagger \kappa),\nonumber 
\,\,
V(H, \xi)= \lambda^{H \xi}_1 (H^\dagger H)(\xi^\dagger \xi),\nonumber 
\,\,
V(\psi, \eta)= \lambda^{\psi \eta}_1 (\psi^\dagger \psi)(\eta^\dagger \eta),\nonumber \\
V(\psi,\chi)=& \lambda^{\psi \chi}_1 (\psi^\dagger \psi)(\chi^\dagger \chi),\nonumber 
\,\,
V(\psi, \kappa)= \lambda^{\psi \kappa}_1 (\psi^\dagger \psi)(\kappa^\dagger \kappa),\nonumber 
\,\,
V(\psi, \xi)= \lambda^{\psi \xi}_1 (\psi^\dagger \psi)(\xi^\dagger \xi),\nonumber \\
V( \psi,\Delta_1)=& \lambda^{ \psi \Delta_1}_1 Tr(\psi^\dagger \psi) (\Delta_1^\dagger \Delta_1),\nonumber 
\quad
V( \psi,\Delta_2)= \lambda^{ \psi \Delta_2}_1 Tr(\psi^\dagger \psi) (\Delta_2^\dagger \Delta_2),\nonumber \\
V( \eta, \chi)=& \lambda^{ \eta \chi}_1 (\eta^\dagger \eta)(\chi^\dagger \chi),\nonumber 
\,\,
V(\eta, \kappa)= \lambda^{\eta \kappa}_1 (\eta^\dagger \eta)(\kappa^\dagger \kappa) \nonumber 
\,\,
V(\eta,\xi)= \lambda^{ \eta \xi}_1(\eta^\dagger \eta) (\xi^\dagger \xi),\nonumber \\
V( \eta,\Delta_1)=& \lambda^{ \eta \Delta_1}_1 Tr(\eta^\dagger \eta)(\Delta_1^\dagger \Delta_1),\nonumber 
\quad
V( \eta,\Delta_2)= \lambda^{ \eta \Delta_2}_1 Tr(\eta^\dagger \eta)(\Delta_2^\dagger \Delta_2),\nonumber\\
V(\xi,\Delta_2)=& \lambda^{ \xi \Delta_2}_1 Tr(\xi^\dagger \xi)(\Delta_2^\dagger \Delta_2),\nonumber 
\quad
V(\Delta_1,\Delta_2)= \lambda^{ \Delta_1 \Delta_2}_1 Tr(\Delta_1^\dagger \Delta_1)(\Delta_2^\dagger \Delta_2),\nonumber\\
V(\chi, \kappa)=& \lambda^{\chi \kappa}_1 (\chi^\dagger \chi)(\kappa^\dagger \kappa),\nonumber 
\,\,
V(\chi, \xi)= \lambda^{\chi \xi}_1 (\chi^\dagger \chi)(\xi^\dagger \xi),\nonumber 
\,\,
V( \kappa,\xi)= \lambda^{ \kappa \xi}_1(\kappa^\dagger \kappa) (\xi^\dagger \xi),\nonumber \\
V(\chi,\Delta_1))=& \lambda^{ \chi \Delta_1}_1 Tr(\chi^\dagger \chi)(\Delta_1^\dagger \Delta_1),\nonumber 
\quad
V(\chi,\Delta_2)= \lambda^{ \chi \Delta_2}_1 Tr(\chi^\dagger \chi)(\Delta_2^\dagger \Delta_2),\nonumber \\
V(\kappa,\Delta_1)=& \lambda^{ \kappa \Delta_1}_1 Tr(\kappa^\dagger \kappa)(\Delta_1^\dagger \Delta_1),\nonumber 
\quad
V(\kappa,\Delta_2)= \lambda^{ \kappa \Delta_2}_1 Tr(\kappa^\dagger \kappa)(\Delta_2^\dagger \Delta_2),\nonumber \\
V(H, \Delta_1) =& 
\lambda_1^{H \Delta_1} (H^\dagger H) \, \text{Tr}(\Delta_1^\dagger \Delta_1)
+ \lambda_2^{H \Delta_1} \left[ (H^\dagger H)_{1'} \, \text{Tr}(\Delta_1^\dagger \Delta_1)_{1''} 
+ (H^\dagger H)_{1''} \, \text{Tr}(\Delta_1^\dagger \Delta_1)_{1'} \right] 
\nonumber \\
&+ \lambda_3^{H \Delta_1} (H^\dagger H)_{3s} \text{Tr}(\Delta_1^\dagger \Delta_1)_{3s} + \lambda_4^{H \Delta_1} \left[ (H^\dagger H)_{3s} \, \text{Tr}(\Delta_1^\dagger \Delta_1)_{3a}
+ (H^\dagger H)_{3a} \, \text{Tr}(\Delta_1^\dagger \right. \nonumber \\
&\left. \Delta_1)_{3s} \right] 
+ \lambda_5^{H \Delta_1} (H^\dagger H)_{3a} \, \text{Tr}(\Delta_1^\dagger \Delta_1)_{3a}
+ Y_1 \eta \left( H^T i \sigma_2 \Delta_1^\dagger H \right), \nonumber \\
V(H, \Delta_2) =& 
\lambda_1^{H \Delta_1} (H^\dagger H) \, \text{Tr}(\Delta_2^\dagger \Delta_2)
+ \lambda_2^{H \Delta_2} \left[ (H^\dagger H)_{1'} \, \text{Tr}(\Delta_2^\dagger \Delta_2)_{1''} 
+ (H^\dagger H)_{1''} \, \text{Tr}(\Delta_2^\dagger \Delta_2)_{1'} \right] 
\nonumber \\
&+ \lambda_3^{H \Delta_2} (H^\dagger H)_{3s} \, 
\text{Tr}(\Delta_2^\dagger \Delta_2)_{3s} + \lambda_4^{H \Delta_2} \left[ (H^\dagger H)_{3s} \, \text{Tr}(\Delta_2^\dagger \Delta_2)_{3a}
+ (H^\dagger H)_{3a} \, \text{Tr}(\Delta_2^\dagger \right.\nonumber \\
&\left. \Delta_2)_{3s} \right] 
+ \lambda_5^{H \Delta_2} (H^\dagger H)_{3a} \, \text{Tr}(\Delta_2^\dagger \Delta_2)_{3a}
+ \frac{Y_2}{\Lambda} \psi \eta \left( H^T i \sigma_2 \Delta_2^\dagger H \right), \nonumber 
\end{align}

Now for the VEV alignments of our choice, the following optimization conditions should be valid,

\begin{align}
\label{optimization}
&\frac{1}{36 \Lambda} 
( 
 \, v_H (15 \sqrt{2} \, v_2 \, v_{\eta} \, v_{\xi} \, Y_2 ) 
+ 8 v_2^2 \Lambda (-9 \lambda_1^{H\Delta_2} + 2 \lambda_3^{H\Delta_2} ) 
 + 2 \Lambda ( -15 v_1 v_{\eta} Y_1 
+ 54 v_1^2 \lambda_1^{H\Delta_1} 
+ 4 v_H^2 \nonumber \\
&(9 \lambda_1^H + 4 \lambda_3^H) 
+ 18 ( 
v_{\eta}^2 \lambda_1^{H\eta} 
+ v_{\kappa}^2 \lambda_1^{H\kappa} + 
v_{\xi}^2 \lambda_1^{H\xi} + 
v_{\chi}^2 \lambda_1^{H\chi} + 
v_{\psi}^2 \lambda_1^{H\psi} 
- 2 \mu_H^2 
) ) 
) = 0,  \nonumber \\
&\frac{1}{36} v_H ( 
 15 v_1 v_{\eta} Y_1 
+ 4 v_1^2 ( 
9 \lambda_2^{H\Delta_1} + 
4 \lambda_3^{H\Delta_1} 
)
+ 4 v_2^2 ( 
-9 \lambda_2^{H\Delta_2} +
2 \lambda_3^{H\Delta_2} 
) 
) = 0, \nonumber \\
&\frac{1}{72 \Lambda} v_H ( 
 15 \sqrt{2} \, v_2 v_{\eta} v_{\xi} Y_2 
+ 30 v_1 v_{\eta} Y_1 \Lambda 
 - 8 v_1^2 \Lambda ( 
9 \lambda_2^{H\Delta_1} + 
4 \lambda_3^{H\Delta_1} 
)
 + 8 v_2^2 \Lambda ( 
9 \lambda_2^{H\Delta_2} - 
2 \lambda_3^{H\Delta_2}  
) 
) \nonumber \\
&= 0, \nonumber \\
&v_1 ( 
 v_{\eta}^2 \lambda_1^{\Delta_1 \eta} 
+ v_{\kappa}^2 \lambda_1^{\Delta_1 \kappa} 
+ v_{\xi}^2 \lambda_1^{\Delta_1 \xi} 
+ v_{\chi}^2 \lambda_1^{\Delta_1 \chi} 
+ v_{\psi}^2 \lambda_1^{\Delta_1 \psi} 
 + v_H^2 ( \lambda_1^{H \Delta_1} + \frac{4}{9} \lambda_3^{H \Delta_1} ) 
+ v_1^2 ( 6 \lambda_1^{\Delta_1} + 2 \lambda_2^{\Delta_1} \nonumber \\
&+ \frac{16}{9} \lambda_3^{\Delta_1} ) 
- 2 \mu_{\Delta_1}^2 
) = 0, \nonumber \\
&v_H^2 ( 
\frac{5}{12} v_{\eta} Y_1 
- v_1 \lambda_1^{H\Delta_1} 
+ \frac{2}{9} v_1 \lambda_3^{H\Delta_1} 
)
- \frac{2}{3} v_1^3 ( 
9 \lambda_1^{\Delta_1} + 4 \lambda_3^{\Delta_1} 
) 
- v_1 ( 
v_{\eta}^2 \lambda_1^{\Delta_1 \eta} 
+ v_{\kappa}^2 \lambda_1^{\Delta_1 \kappa} 
+ v_{\xi}^2 \lambda_1^{\Delta_1 \xi} 
+ \nonumber \\
&v_{\chi}^2 \lambda_1^{\Delta_1 \chi}+ v_{\psi}^2 \lambda_1^{\Delta_1 \psi} 
- 2 \mu_{\Delta_1}^2 
) = 0, \nonumber \\
&\frac{1}{9} v_1 ( 
 v_H^2 ( 9 \lambda_1^{H\Delta_1} - 2 \lambda_3^{H\Delta_1} ) 
+ 6 v_1^2 ( 9 \lambda_1^{\Delta_1} + 4 \lambda_3^{\Delta_1} ) 
+ 9 ( 
v_{\eta}^2 \lambda_1^{\Delta_1 \eta} 
+ v_{\kappa}^2 \lambda_1^{\Delta_1 \kappa} 
+ v_{\xi}^2 \lambda_1^{\Delta_1 \xi} 
+ v_{\chi}^2 \lambda_1^{\Delta_1 \chi} 
+\nonumber \\
& v_{\psi}^2 \lambda_1^{\Delta_1 \psi} - 2 \mu_{\Delta_1}^2 
) 
) = 0, \nonumber \\
&v_H^2 ( 
\frac{5}{12 \sqrt{2} \Lambda} v_{\eta} v_{\xi} Y_2 
- v_2 \lambda_1^{H\Delta_2} 
+ \frac{2}{9} v_2 \lambda_3^{H\Delta_2} 
)
+ v_2^3 ( 
4 \lambda_1^{\Delta_2} 
+ \lambda_2^{\Delta_2} 
+ \frac{4}{3} \lambda_3^{\Delta_2} 
)
- v_2 ( 
v_{\eta}^2 \lambda_1^{\Delta_2 \eta} 
+ v_{\kappa}^2 \lambda_1^{\Delta_2 \kappa} 
\nonumber \\
&+ v_{\xi}^2 \lambda_1^{\Delta_2 \xi} + v_{\chi}^2 \lambda_1^{\Delta_2 \chi} 
+ v_{\psi}^2 \lambda_1^{\Delta_2 \psi} 
- 2 \mu_{\Delta_2}^2 
) = 0, \nonumber \\
&v_H^2 v_2 ( 
\lambda_1^{H\Delta_2} - \frac{2}{9} \lambda_3^{H\Delta_2} 
)
- \frac{1}{3} v_2^3 ( 
12 \lambda_1^{\Delta_2} 
+ 3 \lambda_2^{\Delta_2} 
+ 4 \lambda_3^{\Delta_2} 
)
+ v_2 ( 
v_{\eta}^2 \lambda_1^{\Delta_2 \eta} 
+ v_{\kappa}^2 \lambda_1^{\Delta_2 \kappa} 
+ v_{\xi}^2 \lambda_1^{\Delta_2 \xi} 
+ v_{\chi}^2 \nonumber \\
& \lambda_1^{\Delta_2 \chi} 
+ v_{\psi}^2 \lambda_1^{\Delta_2 \psi} - 2 \mu_{\Delta_2}^2 
) 
= 0, \nonumber 
\end{align}
\begin{align}
&v_H^2 ( 
- \frac{5}{12} v_1 Y_1 
+ \frac{5}{12 \sqrt{2} \Lambda} v_2 v_{\xi} Y_2 
+ v_{\eta} \lambda_1^{H\eta} 
)
+ v_{\eta} ( 
3 v_1^2 \lambda_1^{\Delta_1 \eta} 
+ v_{\kappa}^2 \lambda_1^{\eta \kappa} 
+ v_{\xi}^2 \lambda_1^{\xi \eta} 
- 2 v_2^2 \lambda_1^{\Delta_2 \eta} 
+ v_{\chi}^2 \lambda_1^{\chi \eta} 
\nonumber \\
&+ v_{\psi}^2 \lambda_1^{\psi \eta} 
+ 2 v_{\eta}^2 \lambda^{\eta} - 2 \mu_\eta^2 
) = 0, \nonumber \\
&v_{\kappa} (
v_H^2 \lambda_1^{H\kappa} 
+ 3 v_1^2 \lambda_1^{\Delta_1 \kappa} 
+ v_{\eta}^2 \lambda_1^{\eta \kappa} 
+ v_{\xi}^2 \lambda_1^{\xi \kappa} 
- 2 v_2^2 \lambda_1^{\Delta_2 \kappa} 
+ v_{\chi}^2 \lambda_1^{\chi \kappa} 
+ v_{\psi}^2 \lambda_1^{\psi \kappa} 
+ 2 v_{\kappa}^2 \lambda^{\kappa} 
- 2 \mu_\kappa^2 
) = 0,\nonumber \\
&v_H^2 ( 
\frac{5}{12 \sqrt{2} \Lambda} v_2 v_{\eta} Y_2 
+ v_{\xi} \lambda_1^{H\xi} 
) 
+ v_{\xi} ( 
3 v_1^2 \lambda_1^{\Delta_1 \xi} 
+ v_{\eta}^2 \lambda_1^{\xi \eta} 
+ v_{\kappa}^2 \lambda_1^{\xi \kappa} 
- 2 v_2^2 \lambda_1^{\Delta_2 \xi} 
+ v_{\chi}^2 \lambda_1^{\chi \xi} 
+ v_{\psi}^2 \lambda_1^{\psi \xi} 
+ \nonumber \\
&2 v_{\xi}^2 \lambda^{\xi} 
- 2 \mu_\xi^2 
) = 0,\nonumber \\
&v_{\psi} (
v_H^2 \lambda_1^{H\psi} 
+ 3 v_1^2 \lambda_1^{\Delta_1 \psi} 
- 2 v_2^2 \lambda_1^{\Delta_2 \psi} 
+ v_{\chi}^2 \lambda_1^{\chi \psi} 
+ v_{\eta}^2 \lambda_1^{\psi \eta} 
+ v_{\kappa}^2 \lambda_1^{\psi \kappa} 
+ v_{\xi}^2 \lambda_1^{\psi \xi} 
+ 2 v_{\psi}^2 \lambda^{\psi} 
- 2 \mu_\psi^2 
) = 0,\nonumber \\
&v_{\chi} (
v_H^2 \lambda_1^{H\chi} 
+ 3 v_1^2 \lambda_1^{\Delta_1 \chi} 
- 2 v_2^2 \lambda_1^{\Delta_2 \chi} 
+ v_{\eta}^2 \lambda_1^{\chi \eta} 
+ v_{\kappa}^2 \lambda_1^{\chi \kappa} 
+ v_{\xi}^2 \lambda_1^{\chi \xi} 
+ v_{\psi}^2 \lambda_1^{\chi \psi} 
+ 2 v_{\chi}^2 \lambda^{\chi} 
- 2 \mu_\chi^2 
) = 0. \nonumber
\end{align}

\bibliographystyle{spphys}
\bibliography{paper2.bib}

@article{Davis:1968cp,
    author = "Davis, Jr., Raymond and Harmer, Don S. and Hoffman, Kenneth C.",
    title = "{Search for neutrinos from the sun}",
    doi = "10.1103/PhysRevLett.20.1205",
    journal = "Phys. Rev. Lett.",
    volume = "20",
    pages = "1205--1209",
    year = "1968"
}

@article{SNO:2001kpb,
    author = "Ahmad, Q. R. and others",
    collaboration = "SNO",
    title = "{Measurement of the rate of $\nu_e+d \to p+p+e^-$ interactions produced by $^8$B solar neutrinos at the Sudbury Neutrino Observatory}",
    eprint = "nucl-ex/0106015",
    archivePrefix = "arXiv",
    reportNumber = "UPR-0240E",
    doi = "10.1103/PhysRevLett.87.071301",
    journal = "Phys. Rev. Lett.",
    volume = "87",
    pages = "071301",
    year = "2001"
}

@article{Super-Kamiokande:2001ljr,
    author = "Fukuda, S. and others",
    collaboration = "Super-Kamiokande",
    title = "{Solar B-8 and hep neutrino measurements from 1258 days of Super-Kamiokande data}",
    eprint = "hep-ex/0103032",
    archivePrefix = "arXiv",
    doi = "10.1103/PhysRevLett.86.5651",
    journal = "Phys. Rev. Lett.",
    volume = "86",
    pages = "5651--5655",
    year = "2001"
}

@article{Bionta:1987qt,
    author = "Bionta, R. M. and others",
    title = "{Observation of a Neutrino Burst in Coincidence with Supernova SN 1987a in the Large Magellanic Cloud}",
    reportNumber = "UCI-NEUTRINO-87-10",
    doi = "10.1103/PhysRevLett.58.1494",
    journal = "Phys. Rev. Lett.",
    volume = "58",
    pages = "1494",
    year = "1987"
}

@article{Super-Kamiokande:1998kpq,
    author = "Fukuda, Y. and others",
    collaboration = "Super-Kamiokande",
    title = "{Evidence for oscillation of atmospheric neutrinos}",
    eprint = "hep-ex/9807003",
    archivePrefix = "arXiv",
    reportNumber = "BU-98-17, ICRR-REPORT-422-98-18, UCI-98-8, KEK-PREPRINT-98-95, LSU-HEPA-5-98, UMD-98-003, SBHEP-98-5, TKU-PAP-98-06, TIT-HPE-98-09",
    doi = "10.1103/PhysRevLett.81.1562",
    journal = "Phys. Rev. Lett.",
    volume = "81",
    pages = "1562--1567",
    year = "1998"
}

@article{KamLAND:2002uet,
    author = "Eguchi, K. and others",
    collaboration = "KamLAND",
    title = "{First results from KamLAND: Evidence for reactor anti-neutrino disappearance}",
    eprint = "hep-ex/0212021",
    archivePrefix = "arXiv",
    doi = "10.1103/PhysRevLett.90.021802",
    journal = "Phys. Rev. Lett.",
    volume = "90",
    pages = "021802",
    year = "2003"
}

@article{K2K:2002icj,
    author = "Ahn, M. H. and others",
    collaboration = "K2K",
    title = "{Indications of neutrino oscillation in a 250 km long baseline experiment}",
    eprint = "hep-ex/0212007",
    archivePrefix = "arXiv",
    doi = "10.1103/PhysRevLett.90.041801",
    journal = "Phys. Rev. Lett.",
    volume = "90",
    pages = "041801",
    year = "2003"
}

@article{T2K:2013ppw,
    author = "Abe, K. and others",
    collaboration = "T2K",
    title = "{Observation of Electron Neutrino Appearance in a Muon Neutrino Beam}",
    eprint = "1311.4750",
    archivePrefix = "arXiv",
    primaryClass = "hep-ex",
    doi = "10.1103/PhysRevLett.112.061802",
    journal = "Phys. Rev. Lett.",
    volume = "112",
    pages = "061802",
    year = "2014"
}

@article{Kamiokande-II:1989hkh,
    author = "Hirata, K. S. and others",
    collaboration = "Kamiokande-II",
    title = "{Observation of B-8 Solar Neutrinos in the Kamiokande-II Detector}",
    reportNumber = "ICR-188-89-5, KEK-Preprint-89-2, KOBE-AP-89-04, OULNS-89-01, UPR-0165E",
    doi = "10.1103/PhysRevLett.63.16",
    journal = "Phys. Rev. Lett.",
    volume = "63",
    pages = "16",
    year = "1989"
}

@article{Esteban:2024eli,
    author = "Esteban, Ivan and Gonzalez-Garcia, M. C. and Maltoni, Michele and Martinez-Soler, Ivan and Pinheiro, Jo\~ao Paulo and Schwetz, Thomas",
    title = "{NuFit-6.0: updated global analysis of three-flavor neutrino oscillations}",
    eprint = "2410.05380",
    archivePrefix = "arXiv",
    primaryClass = "hep-ph",
    reportNumber = "IFT-UAM/CSIC-24-140, YITP-SB-2024-24, IPPP/24/64, IPPP/24/64, IFT-UAM/CSIC-24-140, YITP-SB-2024-24",
    doi = "10.1007/JHEP12(2024)216",
    journal = "JHEP",
    volume = "12",
    pages = "216",
    year = "2024"
}

@article{ParticleDataGroup:2022pth,
    author = "Workman, R. L. and others",
    collaboration = "Particle Data Group",
    title = "{Review of Particle Physics}",
    doi = "10.1093/ptep/ptac097",
    journal = "PTEP",
    volume = "2022",
    pages = "083C01",
    year = "2022"
}

@article{King:2018kka,
    author = "King, Stephen F. and Nishi, Celso C.",
    title = "{Mu-tau symmetry and the Littlest Seesaw}",
    eprint = "1807.00023",
    archivePrefix = "arXiv",
    primaryClass = "hep-ph",
    doi = "10.1016/j.physletb.2018.08.056",
    journal = "Phys. Lett. B",
    volume = "785",
    pages = "391--398",
    year = "2018"
}

@article{Xing:2015fdg,
    author = "Xing, Zhi-zhong and Zhao, Zhen-hua",
    title = "{A review of \ensuremath{\mu}-\ensuremath{\tau} flavor symmetry in neutrino physics}",
    eprint = "1512.04207",
    archivePrefix = "arXiv",
    primaryClass = "hep-ph",
    doi = "10.1088/0034-4885/79/7/076201",
    journal = "Rept. Prog. Phys.",
    volume = "79",
    number = "7",
    pages = "076201",
    year = "2016"
}

@article{Altarelli:2005yx,
    author = "Altarelli, Guido and Feruglio, Ferruccio",
    title = "{Tri-bimaximal neutrino mixing, A(4) and the modular symmetry}",
    eprint = "hep-ph/0512103",
    archivePrefix = "arXiv",
    reportNumber = "CERN-PH-TH-2005-226",
    doi = "10.1016/j.nuclphysb.2006.02.015",
    journal = "Nucl. Phys. B",
    volume = "741",
    pages = "215--235",
    year = "2006"
}

@article{Ludl:2014axa,
    author = "Ludl, Patrick Otto and Grimus, Walter",
    title = "{A complete survey of texture zeros in the lepton mass matrices}",
    eprint = "1406.3546",
    archivePrefix = "arXiv",
    primaryClass = "hep-ph",
    reportNumber = "UWTHPH-2014-13",
    doi = "10.1007/JHEP07(2014)090",
    journal = "JHEP",
    volume = "07",
    pages = "090",
    year = "2014",
    note = "[Erratum: JHEP 10, 126 (2014)]"
}

@article{Kalita:2015tda,
    author = "Kalita, Rupam and Borah, Debasish",
    title = "{Hybrid Textures of Neutrino Mass Matrix under the Lamppost of Latest Neutrino and Cosmology Data}",
    eprint = "1509.02728",
    archivePrefix = "arXiv",
    primaryClass = "hep-ph",
    doi = "10.1142/S0217751X16500081",
    journal = "Int. J. Mod. Phys. A",
    volume = "31",
    number = "06",
    pages = "1650008",
    year = "2016"
}

@article{Dey:2022qpu,
    author = "Dey, Manash and Chakraborty, Pralay and Roy, Subhankar",
    title = "{The {\ensuremath{\mu}}-{\ensuremath{\tau}} mixed symmetry and neutrino mass matrix}",
    eprint = "2211.01314",
    archivePrefix = "arXiv",
    primaryClass = "hep-ph",
    doi = "10.1016/j.physletb.2023.137767",
    journal = "Phys. Lett. B",
    volume = "839",
    pages = "137767",
    year = "2023",
    note = "[Erratum: Phys.Lett.B 852, 138602 (2024)]"
}

@article{Chakraborty:2023msb,
    author = "Chakraborty, Pralay and Roy, Subhankar",
    title = "{The other variants of mixed {\ensuremath{\mu}}-{\ensuremath{\tau}} symmetry}",
    eprint = "2304.06737",
    archivePrefix = "arXiv",
    primaryClass = "hep-ph",
    doi = "10.1016/j.nuclphysb.2023.116252",
    journal = "Nucl. Phys. B",
    volume = "992",
    pages = "116252",
    year = "2023"
}

@article{Singh:2018bap,
    author = "Singh, Madan",
    title = "{Investigating the hybrid textures of neutrino mass matrix for near maximal atmospheric neutrino mixing}",
    eprint = "1803.10754",
    archivePrefix = "arXiv",
    primaryClass = "hep-ph",
    doi = "10.1155/2018/5806743",
    journal = "Adv. High Energy Phys.",
    volume = "2018",
    pages = "5806743",
    year = "2018"
}

@article{Harrison:2002er,
    author = "Harrison, P. F. and Perkins, D. H. and Scott, W. G.",
    title = "{Tri-bimaximal mixing and the neutrino oscillation data}",
    eprint = "hep-ph/0202074",
    archivePrefix = "arXiv",
    reportNumber = "RAL-TR-2002-002",
    doi = "10.1016/S0370-2693(02)01336-9",
    journal = "Phys. Lett. B",
    volume = "530",
    pages = "167",
    year = "2002"
}

@article{Shimizu:2011xg,
    author = "Shimizu, Yusuke and Tanimoto, Morimitsu and Watanabe, Atsushi",
    title = "{Breaking Tri-bimaximal Mixing and Large $\theta_{13}$}",
    eprint = "1105.2929",
    archivePrefix = "arXiv",
    primaryClass = "hep-ph",
    doi = "10.1143/PTP.126.81",
    journal = "Prog. Theor. Phys.",
    volume = "126",
    pages = "81--90",
    year = "2011"
}

@article{Xing:2002sw,
    author = "Xing, Zhi-zhong",
    title = "{Nearly tri bimaximal neutrino mixing and CP violation}",
    eprint = "hep-ph/0204049",
    archivePrefix = "arXiv",
    reportNumber = "BIHEP-TH-2002-13",
    doi = "10.1016/S0370-2693(02)01649-0",
    journal = "Phys. Lett. B",
    volume = "533",
    pages = "85--93",
    year = "2002"
}

@article{Xing:2006ms,
    author = "Xing, Zhi-zhong and Zhou, Shun",
    title = "{Tri-bimaximal Neutrino Mixing and Flavor-dependent Resonant Leptogenesis}",
    eprint = "hep-ph/0607302",
    archivePrefix = "arXiv",
    doi = "10.1016/j.physletb.2007.08.009",
    journal = "Phys. Lett. B",
    volume = "653",
    pages = "278--287",
    year = "2007"
}

@article{Adhikary:2006jx,
    author = "Adhikary, Biswajit and Ghosal, Ambar",
    title = "{Constraining it CP violation in a softly broken A(4) symmetric Model}",
    eprint = "hep-ph/0609193",
    archivePrefix = "arXiv",
    reportNumber = "SINP-TNP-06-24",
    doi = "10.1103/PhysRevD.75.073020",
    journal = "Phys. Rev. D",
    volume = "75",
    pages = "073020",
    year = "2007"
}

@article{King:2007pr,
    author = "King, S. F.",
    title = "{Parametrizing the lepton mixing matrix in terms of deviations from tri-bimaximal mixing}",
    eprint = "0710.0530",
    archivePrefix = "arXiv",
    primaryClass = "hep-ph",
    doi = "10.1016/j.physletb.2007.10.078",
    journal = "Phys. Lett. B",
    volume = "659",
    pages = "244--251",
    year = "2008"
}

@article{Brahmachari:2008fn,
    author = "Brahmachari, Biswajoy and Choubey, Sandhya and Mitra, Manimala",
    title = "{The A(4) flavor symmetry and neutrino phenomenology}",
    eprint = "0801.3554",
    archivePrefix = "arXiv",
    primaryClass = "hep-ph",
    reportNumber = "HRI-P-08-01-001, KEK-TH-1220",
    doi = "10.1103/PhysRevD.77.119901",
    journal = "Phys. Rev. D",
    volume = "77",
    pages = "073008",
    year = "2008",
    note = "[Erratum: Phys.Rev.D 77, 119901 (2008)]"
}

@article{Adhikary:2008au,
    author = "Adhikary, Biswajit and Ghosal, Ambar",
    title = "{Nonzero U(e3), CP violation and leptogenesis in a see-saw type softly broken A(4) symmetric model}",
    eprint = "0803.3582",
    archivePrefix = "arXiv",
    primaryClass = "hep-ph",
    doi = "10.1103/PhysRevD.78.073007",
    journal = "Phys. Rev. D",
    volume = "78",
    pages = "073007",
    year = "2008"
}

@article{Hirsch:2008mg,
    author = "Hirsch, M. and Morisi, S. and Valle, J. W. F.",
    title = "{Modelling tri-bimaximal neutrino mixing}",
    eprint = "0810.0121",
    archivePrefix = "arXiv",
    primaryClass = "hep-ph",
    reportNumber = "IFIC-08-50",
    doi = "10.1103/PhysRevD.79.016001",
    journal = "Phys. Rev. D",
    volume = "79",
    pages = "016001",
    year = "2009"
}

@article{Morisi:2009qa,
    author = "Morisi, Stefano",
    title = "{Tri-Bimaximal lepton mixing with A(4) x Z(2)**3}",
    eprint = "0901.1080",
    archivePrefix = "arXiv",
    primaryClass = "hep-ph",
    doi = "10.1103/PhysRevD.79.033008",
    journal = "Phys. Rev. D",
    volume = "79",
    pages = "033008",
    year = "2009"
}

@article{Hayakawa:2009va,
    author = "Hayakawa, Atsushi and Ishimori, Hajime and Shimizu, Yusuke and Tanimoto, Morimitsu",
    title = "{Deviation from tri-bimaximal mixing and flavor symmetry breaking in a seesaw type A(4) model}",
    eprint = "0904.3820",
    archivePrefix = "arXiv",
    primaryClass = "hep-ph",
    doi = "10.1016/j.physletb.2009.09.012",
    journal = "Phys. Lett. B",
    volume = "680",
    pages = "334--342",
    year = "2009"
}

@article{Goswami:2009yy,
    author = "Goswami, Srubabati and Petcov, Serguey T. and Ray, Shamayita and Rodejohann, Werner",
    title = "{Large |U(e3)| and Tri-bimaximal Mixing}",
    eprint = "0907.2869",
    archivePrefix = "arXiv",
    primaryClass = "hep-ph",
    reportNumber = "SISSA-40-2009-EP, TIFR-TH-09-20",
    doi = "10.1103/PhysRevD.80.053013",
    journal = "Phys. Rev. D",
    volume = "80",
    pages = "053013",
    year = "2009"
}

@article{Barry:2010zk,
    author = "Barry, James and Rodejohann, Werner",
    title = "{Deviations from tribimaximal mixing due to the vacuum expectation value misalignment in $A_4$ models}",
    eprint = "1003.2385",
    archivePrefix = "arXiv",
    primaryClass = "hep-ph",
    doi = "10.1103/PhysRevD.81.119901",
    journal = "Phys. Rev. D",
    volume = "81",
    pages = "093002",
    year = "2010",
    note = "[Erratum: Phys.Rev.D 81, 119901 (2010)]"
}

@article{Albright:2010ap,
    author = "Albright, Carl H. and Dueck, Alexander and Rodejohann, Werner",
    title = "{Possible Alternatives to Tri-bimaximal Mixing}",
    eprint = "1004.2798",
    archivePrefix = "arXiv",
    primaryClass = "hep-ph",
    reportNumber = "FERMILAB-PUB-10-079-T",
    doi = "10.1140/epjc/s10052-010-1492-2",
    journal = "Eur. Phys. J. C",
    volume = "70",
    pages = "1099--1110",
    year = "2010"
}

@article{Ishimori:2010fs,
    author = "Ishimori, Hajime and Shimizu, Yusuke and Tanimoto, Morimitsu and Watanabe, Atsushi",
    title = "{Neutrino masses and mixing from $S_{4}$ flavor twisting}",
    eprint = "1010.3805",
    archivePrefix = "arXiv",
    primaryClass = "hep-ph",
    doi = "10.1103/PhysRevD.83.033004",
    journal = "Phys. Rev. D",
    volume = "83",
    pages = "033004",
    year = "2011"
}

@article{King:2011ab,
    author = "King, Stephen F. and Luhn, Christoph",
    title = "{A4 models of tri-bimaximal-reactor mixing}",
    eprint = "1112.1959",
    archivePrefix = "arXiv",
    primaryClass = "hep-ph",
    reportNumber = "IPPP-11-83, DCPT-11-166",
    doi = "10.1007/JHEP03(2012)036",
    journal = "JHEP",
    volume = "03",
    pages = "036",
    year = "2012"
}

@article{Antusch:2011qg,
    author = "Antusch, Stefan and Maurer, Vinzenz",
    title = "{Large neutrino mixing angle $\theta_{13}${\textasciicircum}{MNS} and quark-lepton mass ratios in unified flavour models}",
    eprint = "1107.3728",
    archivePrefix = "arXiv",
    primaryClass = "hep-ph",
    doi = "10.1103/PhysRevD.84.117301",
    journal = "Phys. Rev. D",
    volume = "84",
    pages = "117301",
    year = "2011"
}

@article{Ahn:2012tv,
    author = "Ahn, Y. H. and Kang, Sin Kyu",
    title = "{Non-zero $\theta_{13}$ and CP violation in a model with $A_4$ flavor symmetry}",
    eprint = "1203.4185",
    archivePrefix = "arXiv",
    primaryClass = "hep-ph",
    reportNumber = "KIAS-P12022",
    doi = "10.1103/PhysRevD.86.093003",
    journal = "Phys. Rev. D",
    volume = "86",
    pages = "093003",
    year = "2012"
}

@article{Ishimori:2012fg,
    author = "Ishimori, Hajime and Ma, Ernest",
    title = "{New Simple $A_4$ Neutrino Model for Nonzero $\theta_{13}$ and Large $\delta_{CP}$}",
    eprint = "1205.0075",
    archivePrefix = "arXiv",
    primaryClass = "hep-ph",
    reportNumber = "UCRHEP-T520",
    doi = "10.1103/PhysRevD.86.045030",
    journal = "Phys. Rev. D",
    volume = "86",
    pages = "045030",
    year = "2012"
}

@article{Rodejohann:2012cf,
    author = "Rodejohann, Werner and Zhang, He",
    title = "{Simple two Parameter Description of Lepton Mixing}",
    eprint = "1207.1225",
    archivePrefix = "arXiv",
    primaryClass = "hep-ph",
    doi = "10.1103/PhysRevD.86.093008",
    journal = "Phys. Rev. D",
    volume = "86",
    pages = "093008",
    year = "2012"
}

@article{Hagedorn:2012ut,
    author = "Hagedorn, Claudia and King, Stephen F. and Luhn, Christoph",
    title = "{SUSY S$_{4} \times$ SU(5) revisited}",
    eprint = "1205.3114",
    archivePrefix = "arXiv",
    primaryClass = "hep-ph",
    reportNumber = "DFPD-12-TH-2, IPPP-12-27, DCPT-12-54",
    doi = "10.1016/j.physletb.2012.09.026",
    journal = "Phys. Lett. B",
    volume = "717",
    pages = "207--213",
    year = "2012"
}

@article{King:2013eh,
    author = "King, Stephen F. and Luhn, Christoph",
    title = "{Neutrino Mass and Mixing with Discrete Symmetry}",
    eprint = "1301.1340",
    archivePrefix = "arXiv",
    primaryClass = "hep-ph",
    reportNumber = "IPPP-12-100, DCPT-12-200",
    doi = "10.1088/0034-4885/76/5/056201",
    journal = "Rept. Prog. Phys.",
    volume = "76",
    pages = "056201",
    year = "2013"
}

@article{King:2002nf,
    author = "King, S. F.",
    title = "{Constructing the large mixing angle MNS matrix in seesaw models with right-handed neutrino dominance}",
    eprint = "hep-ph/0204360",
    archivePrefix = "arXiv",
    reportNumber = "SHEP-02-09",
    doi = "10.1088/1126-6708/2002/09/011",
    journal = "JHEP",
    volume = "09",
    pages = "011",
    year = "2002"
}

@article{Frampton:2004ud,
    author = "Frampton, P. H. and Petcov, S. T. and Rodejohann, W.",
    title = "{On deviations from bimaximal neutrino mixing}",
    eprint = "hep-ph/0401206",
    archivePrefix = "arXiv",
    reportNumber = "SISSA-7-2004-EP",
    doi = "10.1016/j.nuclphysb.2004.03.014",
    journal = "Nucl. Phys. B",
    volume = "687",
    pages = "31--54",
    year = "2004"
}

@article{Altarelli:2004jb,
    author = "Altarelli, Guido and Feruglio, Ferruccio and Masina, Isabella",
    title = "{Can neutrino mixings arise from the charged lepton sector?}",
    eprint = "hep-ph/0402155",
    archivePrefix = "arXiv",
    reportNumber = "DFPD-04-TH-05, CERN-PH-TH-2004-027, ROMA1-TH-1368-04",
    doi = "10.1016/j.nuclphysb.2004.04.012",
    journal = "Nucl. Phys. B",
    volume = "689",
    pages = "157--171",
    year = "2004"
}

@article{Ganguly:2022qxj,
    author = "Ganguly, Joy and Gluza, Janusz and Karmakar, Biswajit",
    title = "{Common origin of \ensuremath{\theta}$_{13}$ and dark matter within the flavor symmetric scoto-seesaw framework}",
    eprint = "2209.08610",
    archivePrefix = "arXiv",
    primaryClass = "hep-ph",
    doi = "10.1007/JHEP11(2022)074",
    journal = "JHEP",
    volume = "11",
    pages = "074",
    year = "2022"
}

@article{Borah:2019ldn,
    author = "Borah, Debasish and Karmakar, Biswajit and Nanda, Dibyendu",
    title = "{Planck scale origin of nonzero $\theta_{13}$ and super-WIMP dark matter}",
    eprint = "1906.02756",
    archivePrefix = "arXiv",
    primaryClass = "hep-ph",
    doi = "10.1103/PhysRevD.100.055014",
    journal = "Phys. Rev. D",
    volume = "100",
    number = "5",
    pages = "055014",
    year = "2019"
}

@article{Rodejohann:2004cg,
    author = "Rodejohann, Werner",
    title = "{Type II seesaw mechanism, deviations from bimaximal neutrino mixing and leptogenesis}",
    eprint = "hep-ph/0403236",
    archivePrefix = "arXiv",
    reportNumber = "SISSA-22-2004-EP",
    doi = "10.1103/PhysRevD.70.073010",
    journal = "Phys. Rev. D",
    volume = "70",
    pages = "073010",
    year = "2004"
}

@article{Borah:2013jia,
    author = "Borah, Debasish",
    title = "{Deviations from Tri-Bimaximal Neutrino Mixing Using Type II Seesaw}",
    eprint = "1307.2426",
    archivePrefix = "arXiv",
    primaryClass = "hep-ph",
    doi = "10.1016/j.nuclphysb.2013.08.024",
    journal = "Nucl. Phys. B",
    volume = "876",
    pages = "575--586",
    year = "2013"
}

@article{King:2011zj,
    author = "King, Stephen F. and Luhn, Christoph",
    title = "{Trimaximal neutrino mixing from vacuum alignment in A4 and S4 models}",
    eprint = "1107.5332",
    archivePrefix = "arXiv",
    primaryClass = "hep-ph",
    reportNumber = "SHEP-11-17",
    doi = "10.1007/JHEP09(2011)042",
    journal = "JHEP",
    volume = "09",
    pages = "042",
    year = "2011"
}

@article{Honda:2008rs,
    author = "Honda, Mizue and Tanimoto, Morimitsu",
    title = "{Deviation from tri-bimaximal neutrino mixing in A(4) flavor symmetry}",
    eprint = "0801.0181",
    archivePrefix = "arXiv",
    primaryClass = "hep-ph",
    doi = "10.1143/PTP.119.583",
    journal = "Prog. Theor. Phys.",
    volume = "119",
    pages = "583--598",
    year = "2008"
}

@article{Ishimori:2010au,
    author = "Ishimori, Hajime and Kobayashi, Tatsuo and Ohki, Hiroshi and Shimizu, Yusuke and Okada, Hiroshi and Tanimoto, Morimitsu",
    title = "{Non-Abelian Discrete Symmetries in Particle Physics}",
    eprint = "1003.3552",
    archivePrefix = "arXiv",
    primaryClass = "hep-th",
    reportNumber = "KUNS-2260",
    doi = "10.1143/PTPS.183.1",
    journal = "Prog. Theor. Phys. Suppl.",
    volume = "183",
    pages = "1--163",
    year = "2010"
}

@article{Maki:1962mu,
    author = "Maki, Ziro and Nakagawa, Masami and Sakata, Shoichi",
    title = "{Remarks on the unified model of elementary particles}",
    doi = "10.1143/PTP.28.870",
    journal = "Prog. Theor. Phys.",
    volume = "28",
    pages = "870--880",
    year = "1962"
}

@article{Planck:2018vyg,
    author = "Aghanim, N. and others",
    collaboration = "Planck",
    title = "{Planck 2018 results. VI. Cosmological parameters}",
    eprint = "1807.06209",
    archivePrefix = "arXiv",
    primaryClass = "astro-ph.CO",
    doi = "10.1051/0004-6361/201833910",
    journal = "Astron. Astrophys.",
    volume = "641",
    pages = "A6",
    year = "2020",
    note = "[Erratum: Astron.Astrophys. 652, C4 (2021)]"
}

@article{DESI:2025zgx,
    author = "Abdul Karim, M. and others",
    collaboration = "DESI",
    title = "{DESI DR2 results. II. Measurements of baryon acoustic oscillations and cosmological constraints}",
    eprint = "2503.14738",
    archivePrefix = "arXiv",
    primaryClass = "astro-ph.CO",
    reportNumber = "FERMILAB-PUB-25-0169-PPD",
    doi = "10.1103/tr6y-kpc6",
    journal = "Phys. Rev. D",
    volume = "112",
    number = "8",
    pages = "083515",
    year = "2025"
}

@article{Schechter:1981bd,
    author = "Schechter, J. and Valle, J. W. F.",
    title = "{Neutrinoless Double beta Decay in SU(2) x U(1) Theories}",
    reportNumber = "SU-4217-213, COO-3533-213",
    doi = "10.1103/PhysRevD.25.2951",
    journal = "Phys. Rev. D",
    volume = "25",
    pages = "2951",
    year = "1982"
}

@article{Ejiri:2020xmm,
    author = "Ejiri, Hiroyasu",
    title = "{Neutrino-Mass Sensitivity and Nuclear Matrix Element for Neutrinoless Double Beta Decay}",
    doi = "10.3390/universe6120225",
    journal = "Universe",
    volume = "6",
    number = "12",
    pages = "225",
    year = "2020"
}

@article{Agostini:2022zub,
    author = "Agostini, Matteo and Benato, Giovanni and Detwiler, Jason A. and Men\'endez, Javier and Vissani, Francesco",
    title = "{Toward the discovery of matter creation with neutrinoless \ensuremath{\beta}\ensuremath{\beta} decay}",
    eprint = "2202.01787",
    archivePrefix = "arXiv",
    primaryClass = "hep-ex",
    doi = "10.1103/RevModPhys.95.025002",
    journal = "Rev. Mod. Phys.",
    volume = "95",
    number = "2",
    pages = "025002",
    year = "2023"
}

@article{CUORE:2019yfd,
    author = "Adams, D. Q. and others",
    collaboration = "CUORE",
    title = "{Improved Limit on Neutrinoless Double-Beta Decay in  $^{130}$Te with CUORE}",
    eprint = "1912.10966",
    archivePrefix = "arXiv",
    primaryClass = "nucl-ex",
    doi = "10.1103/PhysRevLett.124.122501",
    journal = "Phys. Rev. Lett.",
    volume = "124",
    number = "12",
    pages = "122501",
    year = "2020"
}

@article{GERDA:2019ivs,
    author = "Agostini, M. and others",
    collaboration = "GERDA",
    title = "{Probing Majorana neutrinos with double-$\beta$ decay}",
    eprint = "1909.02726",
    archivePrefix = "arXiv",
    primaryClass = "hep-ex",
    doi = "10.1126/science.aav8613",
    journal = "Science",
    volume = "365",
    pages = "1445",
    year = "2019"
}

@article{KamLAND-Zen:2016pfg,
    author = "Gando, A. and others",
    collaboration = "KamLAND-Zen",
    title = "{Search for Majorana Neutrinos near the Inverted Mass Hierarchy Region with KamLAND-Zen}",
    eprint = "1605.02889",
    archivePrefix = "arXiv",
    primaryClass = "hep-ex",
    doi = "10.1103/PhysRevLett.117.082503",
    journal = "Phys. Rev. Lett.",
    volume = "117",
    number = "8",
    pages = "082503",
    year = "2016",
    note = "[Addendum: Phys.Rev.Lett. 117, 109903 (2016)]"
}

@article{SuperNEMO:2021hqx,
    author = "Arnold, R. and others",
    collaboration = "SuperNEMO",
    title = "{Measurement of the distribution of \textasciicircum{}207Bi depositions on calibration sources for SuperNEMO}",
    eprint = "2103.14429",
    archivePrefix = "arXiv",
    primaryClass = "physics.ins-det",
    doi = "10.1088/1748-0221/16/07/T07012",
    journal = "JINST",
    volume = "16",
    number = "07",
    pages = "T07012",
    year = "2021"
}

@article{CUORE:2018ncg,
    author = "Alduino, C. and others",
    collaboration = "CUORE",
    title = "{Double-beta decay of $^{130}\hbox {Te}$ to the first $0^+$ excited state of $^{130}\hbox {Xe}$ with CUORE-0}",
    eprint = "1811.10363",
    archivePrefix = "arXiv",
    primaryClass = "nucl-ex",
    doi = "10.1140/epjc/s10052-019-7275-5",
    journal = "Eur. Phys. J. C",
    volume = "79",
    number = "9",
    pages = "795",
    year = "2019"
}

@article{Chakraborty:2024eki,
    author = "Chakraborty, Pralay and Dey, Manash and Karmakar, Biswajit and Roy, Subhankar",
    title = "{Neutrino mixing from a fresh perspective}",
    eprint = "2405.10353",
    archivePrefix = "arXiv",
    primaryClass = "hep-ph",
    doi = "10.1016/j.physletb.2024.139020",
    journal = "Phys. Lett. B",
    volume = "858",
    pages = "139020",
    year = "2024"
}

@article{Vien:2025fiu,
    author = "Vien, V. V.",
    title = "{Neutrino mass and mixing from a novel scenario with $A_4$ symmetry}",
    doi = "10.1140/epjp/s13360-025-06346-5",
    journal = "Eur. Phys. J. Plus",
    volume = "140",
    number = "5",
    pages = "399",
    year = "2025"
}

@article{Chakraborty:2024hhq,
    author = "Chakraborty, Pralay and Roy, Subhankar",
    title = "{A unique neutrino mass matrix texture under exponential parametrization}",
    eprint = "2407.02550",
    archivePrefix = "arXiv",
    primaryClass = "hep-ph",
    doi = "10.1007/JHEP07(2025)246",
    journal = "JHEP",
    volume = "07",
    pages = "246",
    year = "2025"
}

@article{Chakraborty:2025juy,
    author = "Chakraborty, Pralay and Goswami, Sagar Tirtha and Roy, Subhankar",
    title = "{A Highly Predictive Neutrino Model: The Step Toward Precision}",
    eprint = "2508.07837",
    archivePrefix = "arXiv",
    primaryClass = "hep-ph",
    month = "8",
    year = "2025"
}

@article{Capozzi:2025ovi,
    author = "Capozzi, Francesco and Lisi, Eligio and Marcone, Francesco and Marrone, Antonio and Palazzo, Antonio",
    title = "{Updated bounds on the (1,2) neutrino oscillation parameters after first JUNO results}",
    eprint = "2511.21650",
    archivePrefix = "arXiv",
    primaryClass = "hep-ph",
    month = "11",
    year = "2025"
}

@article{JUNO:2025gmd,
    author = "Abusleme, Angel and others",
    collaboration = "JUNO",
    title = "{First measurement of reactor neutrino oscillations at JUNO}",
    eprint = "2511.14593",
    archivePrefix = "arXiv",
    primaryClass = "hep-ex",
    month = "11",
    year = "2025"
}

@article{Esteban:2026phq,
    author = "Esteban, Ivan and Gonzalez-Garcia, M. C. and Maltoni, Michele and Martinez-Soler, Ivan and Pinheiro, Joao Paulo and Schwetz, Thomas",
    title = "{Lessons from the first JUNO results}",
    eprint = "2601.09791",
    archivePrefix = "arXiv",
    primaryClass = "hep-ph",
    reportNumber = "IFT-UAM/CSIC-26-3, IPPP/26/03, YITP-SB-2026-02",
    month = "1",
    year = "2026"
}

\end{document}